\documentclass{aa}  
\usepackage{caption}
\usepackage[varg]{txfonts}
\usepackage[dvipsnames]{xcolor}
\usepackage{graphicx}
\usepackage{amssymb}
\usepackage{pifont}
\usepackage{soul}
\usepackage{placeins}
\newcommand{\cmark}{\ding{51}}%
\newcommand{\xmark}{\ding{55}}%

\usepackage{txfonts}
\usepackage{hyperref}

\begin{document} 

\renewcommand{\arraystretch}{0.5}

\title{The OTELO survey}
\subtitle{III. Demography, morphology, IR luminosity and environment of AGN hosts}
\author{\ Marina Ram\'on-P\'erez\inst{1,2}
\and \'Angel Bongiovanni\inst{1,2,3}
\and Ana Mar\'ia P\'erez Garc\'ia\inst{4,3}
\and Jordi Cepa\inst{1,2,3}
\and Jakub Nadolny\inst{1,2}
\and Irene Pintos-Castro\inst{5}
\and Maritza A. Lara-L\'opez\inst{6}
\and Emilio J. Alfaro\inst{7}
\and H\'ector O. Casta\~neda\inst{8}
\and Miguel Cervi\~no\inst{4,1,7}
\and Jos\'e A. de Diego\inst{9}
\and Mirian Fern\'andez-Lorenzo\inst{7}
\and Jes\'us Gallego\inst{10}
\and J. Jes\'us Gonz\'alez\inst{9}
\and J. Ignacio Gonz\'alez-Serrano\inst{11,3}
\and Iv\'an Oteo G\'omez\inst{12,13}
\and Ricardo P\'erez Mart\'inez\inst{14,3}
\and Mirjana Povi\'c\inst{15,7}
\and Miguel S\'anchez-Portal\inst{16,3}
}
\institute{Instituto de Astrof\'isica de Canarias (IAC), E-38200 La Laguna, Tenerife, Spain 
\and Departamento de Astrof\'isica, Universidad de La Laguna (ULL), E-38205 La Laguna, Tenerife, Spain 
\and Asociaci\'on Astrof\'isica para la Promoci\'on de la Investigaci\'on, Instrumentaci\'on y su Desarrollo, ASPID, E-38205 La Laguna, Tenerife, Spain 
\and Centro de Astrobiolog\'ia (CSIC/INTA), E-28692, ESAC Campus, Villanueva de la Cañada, Madrid, Spain 
\and Department of Astronomy \& Astrophysics, University of Toronto, Canada 
\and DARK, Niels Bohr Institute, University of Copenhagen, Lyngbyvej 2, Copenhagen DK-2100, Denmark 
\and Instituto de Astrof\'isica de Andaluc\'ia, CSIC, E-18080, Granada, Spain  
\and Instituto de Astronom\'ia, Universidad Nacional Aut\'onoma de M\'exico, 04510 Ciudad de M\'exico, Mexico 
\and Departamento de F\'isica, Escuela Superior de F\'isica y Matem\'aticas, Instituto Polit\'ecnico Nacional, 07738 Ciudad de M\'exico, Mexico 
\and Departamento de F\'isica de la Tierra y Astrof\'isica \& Instituto de F\'isica de Part\'iculas y del Cosmos (IPARCOS), Facultad de CC. F\'isicas, Universidad Complutense de Madrid, E-28040, Madrid, Spain 
\and Instituto de F\'isica de Cantabria (CSIC-Universidad de Cantabria), E-39005 Santander, Spain 
\and Institute for Astronomy, University of Edinburgh, Royal Observatory, Blackford Hill, Edinburgh, EH9  3HJ, UK 
\and European Southern Observatory, Karl-Schwarzschild-Str. 2, 85748, Garching, Germany  
\and ISDEFE for European Space Astronomy Centre (ESAC)/ESA, P.O. Box 78, E-28690, Villanueva de la Ca\~nada, Madrid, Spain 
\and Ethiopian Space Science and Technology Institute (ESSTI), Entoto Observatory and Research Center (EORC), Astronomy and Astrophysics Research Division, PO Box 33679, Addis Ababa, Ethiopia 
\and Instituto de Radioastronom\'ia Milim\'etrica (IRAM), Av. Divina Pastora 7, N\'ucleo Central, E-18012 Granada, Spain 
}
\date{Received 24 April 2018 / Accepted 25 March 2019}

  \abstract
   {}
   {We take advantage of the capabilities of the OSIRIS Tunable Emission Line Object (OTELO) survey to select and study the AGN population in the field. In particular, we aim to perform an analysis of the properties of these objects, including their demography, morphology, and IR luminosity. Focusing on the population of H$\alpha$ emitters at $z\sim0.4$, we also aim to study the environments of AGN and non-AGN galaxies at that redshift.}
   {We make use of the multiwavelength catalogue of objects in the field compiled by the OTELO survey, unique in terms of minimum flux and equivalent width. We also take advantage of the pseudo-spectra built for each source, which allow the identification of emission lines and the discrimination of different types of objects.}
   {We obtained a sample of 72 AGNs in the field of OTELO, selected with four different methods in the optical, X-rays, and mid-infrared bands. We find that using X-rays is the most efficient way to select AGNs. An analysis was performed on the AGN population of OTELO in order to characterise its members. At $z\sim 0.4$, we find that up to 26\% of our H$\alpha$ emitters are AGNs. At that redshift, AGNs are found in identical environments to non-AGNs, although they represent the most clustered group when compared to passive and star-forming galaxies. The majority of our AGNs at any redshift were classified as late-type galaxies, including a 16\% proportion of irregulars. Another 16\% of AGNs show signs of interactions or mergers. Regarding the infrared luminosity, we are able to recover all the luminous infrared galaxies (LIRGs) in the field of OTELO up to $z\sim 1.6$. We find that the proportion of LIRGs and ultra-luminous infrared galaxies
(ULIRGs) is higher among the AGN population, and that ULIRGs show a higher fraction of AGNs than LIRGs.}  
   {}

\titlerunning{Active Galactic Nuclei in the OTELO survey}
\authorrunning{Ram\'on-P\'erez et al.}
\maketitle

\section{Introduction}
\label{intro}

Galaxies hosting an active galactic nucleus (AGN) show an intense activity in a small, concentrated nuclear region, which makes them much brighter than inactive galaxies of the same Hubble type. Unlike star-forming galaxies, the intense activity of an AGN has a non-stellar origin, although both types of objects display strong emission lines in their spectra. The enormous luminosity of AGNs makes them easily recognisable at great cosmological distances, therefore their study gives us clues about the formation and evolution of galaxies in the Universe \citep{Blandford1990}. Moreover, the analysis of AGN morphologies, their environmental dependencies, and their connection to other relevant astrophysical objects such as luminous and ultra-luminous infrared galaxies (LIRGs and ULIRGs) are also key for our understanding of galaxy evolution. \par

Active galactic nuclei can be selected in a variety of ways based on their different spectral properties. One of the best ways is to perform spectroscopy in the optical or infrared (IR) range, so as to determine if the underlying ionizing continuum is of stellar type or rather follows a power law. The intensity of ultraviolet (UV) and optical emission lines can also be analysed, as proposed by \cite{Baldwin1981} and later by \cite{Veilleux1987}. This method is one of the most reliable ones, but its completeness is difficult to evaluate since it depends on the signal-to-noise ratio (S/N) of the spectra and the redshift of the emission lines \citep{Mushotzky2004}.

When spectroscopy is not available, other selection techniques must be used. Some of these alternative methods imply looking at the galaxy mid-infrared (MIR) colours (\citealt{Lacy2004}, \citealt{Stern2005}, \citealt{Donley2012}). One of the main features of AGNs is the power-law continuum that generally dominates their spectrum from UV to $\sim$5 $\mu$m. On the contrary, star-forming galaxies exhibit a blackbody-shaped continuum due to their stellar populations in this range, with a peak around $\sim$1.6 $\mu$m. As a consequence, AGNs tend to be redder than normal galaxies in the MIR. By using IR colours, one can obtain information about the underlying continuum in a spectrum and detect objects whose spectral energy distribution (SED) does not decline in the red side of the stellar peak. The great advantage of MIR selection of AGNs is that it permits to detect even those objects obscured by interstellar gas or by dust that cannot be seen in X-rays or in the optical. However, when compared to other bands, images in IR may sometimes suffer from poorer spatial resolution. Another drawback is that at intermediate luminosities, AGN selection in the IR seems to be biased towards unobscured AGNs \citep{Messias2014}. \par 

Other AGN-selection techniques focus on the X-ray emission, as it is a very good indicator of nuclear activity in galaxies \citep{Mushotzky2004}. In fact, AGNs are believed to be the prevailing astronomical objects contributing to the cosmic X-ray background \citep{DellaCeca2004}. In the surveys carried out with the Chandra and XMM-Newton spatial observatories for instance, the majority of the extragalactic X-ray sources that were found were AGNs \citep{Brandt2004}. The strong X-ray emission of those objects is produced in the central regions of the accretion disc surrounding the black hole.\par 
Due to the diversity of AGN types, a specific technique may correctly select an AGN population while missing others. For instance, a selection based on X-ray or optical emission can miss the population of obscured (either by interstellar gas or by dust) AGNs, unlike an IR-photometry-based method. On the other hand, X-ray emission is a powerful tool to select low-luminosity AGNs or AGNs hiding behind larger hydrogen column densities than those found by optical methods. That is why a multiwavelength approach is preferable in order to obtain reliable unbiased AGN datasets.\par \bigskip

In recent years, the use of tunable filters (TFs) in large telescopes has begun to stand out as an efficient way of obtaining low-resolution spectroscopy of a large number of sources simultaneously, and also exploring the sky at deeper magnitudes \citep[see][and references therein]{OTELO1}. This technique is particularly useful for the detection of emitting objects even at high redshifts. The OSIRIS Tunable Emission Line Object project, OTELO\footnote{\tt http://research.iac.es/proyecto/otelo}, is an ambitious  emission-line survey which makes use of the red TF of the OSIRIS instrument \citep{cepa2003}, installed in the 10.4 m Gran Telescopio Canarias (GTC), currently the largest fully steerable optical reflecting telescope in the world \citep{Alvarez1998}. OTELO is a blind tomography that samples the spectral range ($9070-9280$ \AA) every 6 \AA\ with a full-width at half maximum (FWHM) of about 12 \AA, allowing for the observation of emission lines in well-defined redshift windows in a selected area of 7.5\arcmin$\times$7.4\arcmin in the Extended Groth Strip. In particular, the H$\alpha$+[NII] lines are observed at $z\sim0.4$. \par 

OTELO is the deepest emission-line survey to date, with unique detection limits in terms of minimum flux and equivalent width \citep[EW;][]{ramonperez2019}. Moreover, a large multiwavelength catalogue of all the sources detected in the field, with data ranging from X-rays to far-infrared (FIR), has been compiled in \cite{OTELO1}. This catalogue contains 11237 entries and is 50\% complete at AB magnitude 26.38. A summary of the available bands in this catalogue is shown in Table \ref{table_catalogues}. \par 

The multiwavelength catalogue of OTELO is a fundamental tool for the identification of all the AGNs in the field using different selection methods in different wavelength bands. Moreover, the pseudo-spectra (PS) directly derived for each source in the field (see \citealt{OTELO1} for a description) allow us to go one step further in the identification of AGNs through their emission-lines. It also permits the identification of all the H$\alpha$ emitters in the field (both AGNs and non-AGNs).\par

In this work, we therefore aim to take advantage of the capabilities of OTELO to study the AGN population in the field. In particular, we aim to perform an analysis of the properties of these objects, including their demography and morphology. Focusing on the population of H$\alpha$ emitters at $z\sim0.4$, an attempt will be made to study the environments of AGNs and non-AGNs at that redshift.\par 

This paper is organised as follows. In section \ref{selectionAGN} the selection of AGNs in the field by different methods is explained. The analysis of AGNs, including their environment, morphology, and the identification of LIRGs and ULIRGs is described in section \ref{analysisAGN}. Finally, section \ref{conclusions} summarises the main conclusions of this work.\par

In this paper we assume a standard $\Lambda$-cold dark matter cosmology with $\Omega_\Lambda$=0.69, $\Omega_{\rm m}$=0.31, and $H_0$=67.8 km s$^{-1}$ Mpc$^{-1}$, as extracted from \cite{planck15}.\par

\section{Selection of AGNs} 
\label{selectionAGN}

We took advantage of the multiwavelength data and PS available for OTELO sources. A summary of the photometric bands included in the catalogue of OTELO can be found in Table \ref{table_catalogues}. Three different techniques were used in order to select AGNs. The first one targets the AGN optical emission and uses a diagnostic diagram to separate them from star-forming galaxies (SFGs). The second one employs the X-ray-to-optical-flux ratio (X/O). Finally, the third one uses MIR colour-colour diagrams.

\begin{table*}[ht]
\vspace*{2mm} 
\caption[]{Available bands in OTELO's multiwavelength catalogue and their corresponding original catalogues (see \citealt{OTELO1} for more details).}       
\vspace*{-5mm}       
\label{table_catalogues}      
\centering    \footnotesize                                
\begin{center}\begin{tabular}{c c c c c}          
\hline   \\     [-1pt]                        
 & Catalogue & Bands  & Reference   \\  [2pt]   
\hline      \\[2pt]                        
    X-rays & Chandra & 0.5–7 keV   & \cite{povic2009}    \\ [4pt] 
    Ultraviolet & Galex &  NUV, FUV    & \cite{Morrissey2007}   \\[4pt] 
    Mid-infrared & {\it Spitzer}/IRAC & 3.6, 4.5, 5.8, \& 8 $\mu$m   & \cite{Barro2011}      \\[4pt] 
    Far-infrared (I) & {\it Spitzer}/MIPS \& {\it Herschel}/PACS  &  24, 100 \& 160 $\mu$m    & \cite{Lutz2011}   \\[4pt] 
    Far-infrared (II)  & {\it Herschel}/SPIRE & 250, 300 \& 500  $\mu$m  & \cite{roseboom2010}  \\[4pt] 
\hline                                             
\end{tabular}\end{center}
\end{table*}
\normalsize

\subsection{AGN at \textit{z} $\sim$ 0.4} 
\label{AGNat04}
The first method to select AGNs benefits from the potential of the OTELO survey to identify emitting objects. The flux excess measured on the PS built for every object in the catalogue of OTELO, or its location in the appropriate colour-magnitude diagram, together with the photometric redshift estimation, allow us to make a preliminary identification of the emission-line source candidates per volume of universe explored (see \citealt{OTELO1}). In this case, a range of $0.3\leq {\rm photo-}z \leq0.5$ was chosen in order to ensure a sample of potential H$\alpha$ candidates that is  as complete as possible. This guess redshift range also takes into account a reported photo-$z$ accuracy $\vert\Delta$z$\vert$/(1+z) $\leq$ 0.2. Subsequently, these candidates were individually examined using a collaborative web-based data visualisation facility that includes a line identifier tool in order to assign scaled likelihood values to the possible line identities, and therefore reliable redshifts. The probability of a given candidate to belong (or not) to the OTELO H$\alpha$ window was then calculated by comparing and weighting the different values of redshift and the corresponding likelihoods assigned to this object.\par

Following this methodology, we obtained a sample of 46 sources whose PSs show an emission compatible with the H$\alpha$+[NII]$\lambda$6584 feature. In our case, the wavelength sampling of the PS would also enable the deblending of the H$\alpha$ and [NII] lines and the measurement of fluxes and EWs \citep{laralopez2011}. This makes it possible to use optical diagnostics aimed to discriminate between AGNs and starburst galaxies. A further description of the selection of H$\alpha$ emitters at $z\sim0.4$ in the field of OTELO is given in \cite{ramonperez2019}.

\subsubsection{Broad-line AGNs}  
\label{BroadLineAGN}

First of all, we selected broad-line AGNs (BLAGNs) from our sample of 46 H$\alpha$ emitters. Broad-line AGNs show permitted lines with widths of thousands of kilometers per second. In comparison, narrow-line AGNs (NLAGNs) have line widths of only a few hundred kilometers per second or less. Their selection is described in Sect.\,\ref{MeasurementHalphaFluxes}.\par 

In order to check the aspect of broad-lines when observed through OSIRIS tunable filters, we first performed a simulation using two real spectra of BLAGNs (Seyferts 1.5 NGC 3516, see \citealt{Arribas1997}, and NGC 4151, see \citealt{Kaspi1996}), following the same methodology as \cite{Sanchez-Portal2015}. We saw that even if the H$\alpha$+[NII] emission was well reproduced in both examples, in the case of NGC 3516 the line is so broad that it becomes diluted after the convolution and the object would fail the automatic test for the detection of emission lines described in \cite{Sanchez-Portal2015}. According to the results of our detection-limit simulations for the case of the H$\alpha$ emission line, an approximate upper detection limit is $\sim$60 \AA\ for the FWHM of the input Gaussian (H$\alpha$), corresponding to $\sim$84 \AA\ at $z=0.398$ and to a width of $\sim$2700 km s$^{-1}$ at that redshift \citep{ramonperez2019}. The situation improves when the lines are not centred in the wavelength window of OTELO. In those cases, when the line appears close to the limiting edges of the spectral window, the pseudo-continuum is more realistically reproduced by the algorithm, favouring the detection. 
 \par 

\begin{figure*}[!htb]
\centering
\includegraphics[width=\textwidth]{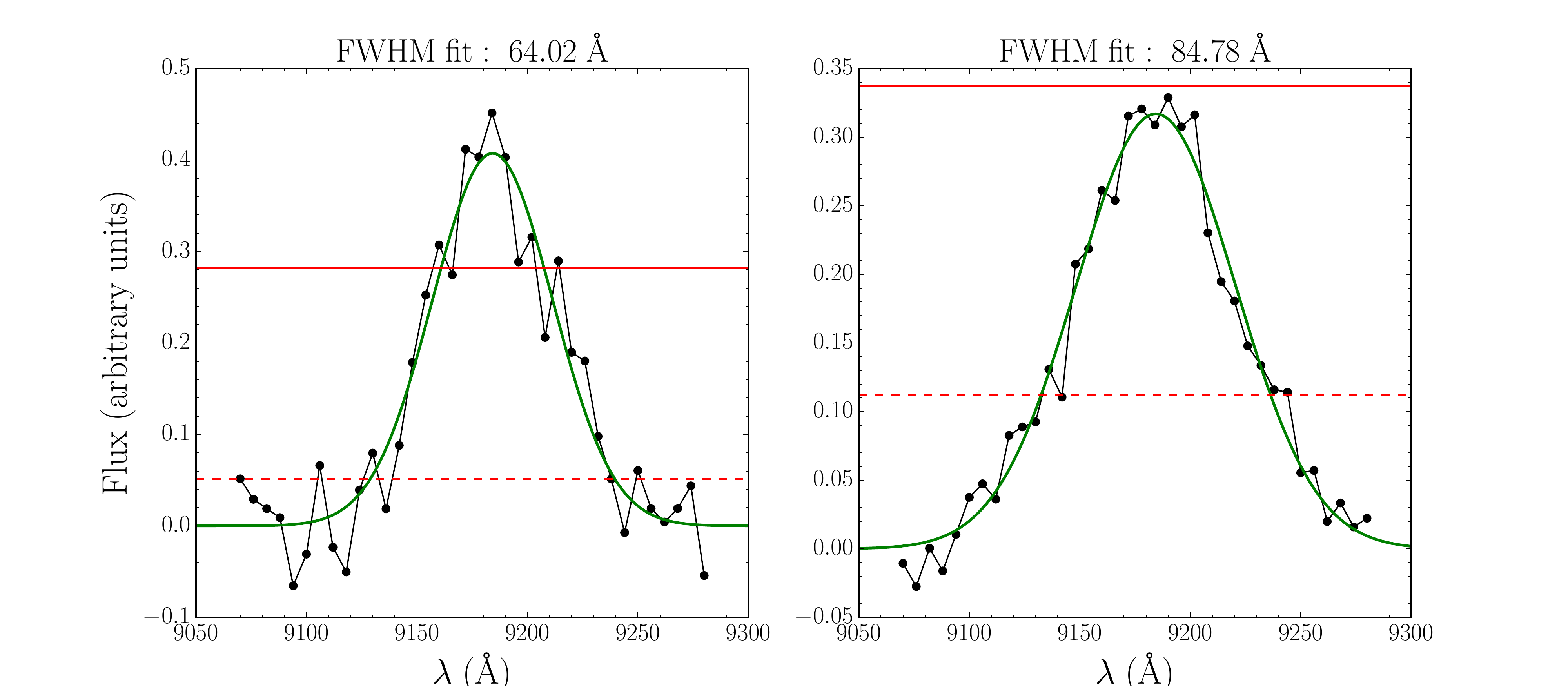}
\caption[{Simulated pseudo-spectra of  H$\alpha$+[NII] lines with different widths}]{Simulated pseudo-spectra (black dots) of  H$\alpha$+[NII] lines with different widths after being convolved to the TF spectral response. The green lines represent the best Gaussian fit to each pseudo-spectrum and the titles describe the corresponding FWHM. The red dashed lines represent the pseudo-continuum, $f_{\rm cont}$, defined as the median value of the pseudo-spectrum points that remain within 2$\sigma$ of the median value of the whole pseudo-spectrum. The red continuous lines represent $f_{\rm cont} + 2\sigma_{\rm cont}$, where $\sigma_{\rm cont}$ is the standard deviation of the pseudo-continuum points. The automatic algorithm used in \cite{ramonperez2019} efficiently detects broad lines as emitting lines for widths up to $\sim$84 \AA. In the left panel, with a FHWM of 64.02 \AA, the broad line is detected. On the contrary, the line in the right panel is so broad (FWHM of 84.78 \AA) that it is not recognised.}
\label{simulations_limit}
\end{figure*}

BLAGN were selected on the basis of two different but equivalent criteria. The first one consisted in fitting a Gaussian to the pseudo-spectrum and determining the FWHM of the fit (as in Fig. \ref{simulations_limit}). After visual inspection to discard incorrect or unclear fits, we selected as BLAGN those objects having a FWHM greater than $\sim$30 \AA\ (corresponding to $\sim$1000 km s$^{-1}$ at $z\sim 0.4$). The second criterion was used whenever the pseudo-spectrum could not be fitted. In those cases, we calculated the number of PS points around the maximum that exceeded half its value. Two different maxima were considered: (i) the real PS maximum and (ii) the closest PS point to the H$\alpha$ line maximum, given the redshift. If at least five points around one of this maximum had a value higher than half maximum, the object was considered a BLAGN. Taking into account that the sampling interval of OTELO is 6 \AA, both this criterion and the previous one are equivalent. In total, six H$\alpha$ emitters were selected as BLAGN by one or both of these criteria. They are shown in  Fig. \ref{BLAGN} of Appendix \ref{appendix_BLAGN}. One of these objects (the last one in the figure) showed a truncated line which prevented the fitting and the analysis. However, it was included in the final sample of BLAGN because its width is comparable to the rest of the objects selected as BLAGN, if we assume a symmetrical line.

\subsubsection{Narrow-line AGNs: measurement of equivalent width and flux of the H$\alpha$ and [NII]$\lambda$6584 emissions}  
\label{MeasurementHalphaFluxes}

The first step in the process of selecting NLAGN was to subtract the continuum of the pseudo-spectra, previously calculated by a linear fit to the data points outside the emission line and visually verified. In some cases, when the continuum fit was not good enough and included part of the emission line, the continuum level was subtracted manually. The H$\alpha$ and [NII] fluxes [$f(\text{H}\alpha)$ and $f(\text{[NII]})$, respectively] were derived following the procedure described in \cite{Sanchez-Portal2015}, that assumes infinitely thin lines. For each object, the redshifted position of both lines in wavelength is known. The fluxes measured in the closest scan slices to these positions ($ f_{\text{H}\alpha}$ and $ f_{\text{[NII]}}$, respectively), correspond to a combination of both line fluxes, such as

\begin{equation}
\begin{aligned}
  f_{\text{H}\alpha} & = T_{\text{H}\alpha}(\text{H}\alpha) f(\text{H}\alpha) +   T_{\text{H}\alpha}(\text{[NII]})f(\text{[NII]}), \\
  f_{\text{[NII]}} & = T_{\text{[NII]}}(\text{H}\alpha) f(\text{H}\alpha) +   T_{\text{[NII]}}(\text{[NII]})f(\text{[NII]}). 
\end{aligned}
\end{equation}

In the previous equations, $T_{\text{`slice'}}(\text{`line'})$  represents the TF transmission of a given slice at a given  wavelength. The real  H$\alpha$ and [NII] fluxes can then be derived from the previous equations as follows.

\begin{equation}
f(\text{H}\alpha) = \frac{ f_{\text{H}\alpha} T_{\text{[NII]}}(\text{[NII]}) - f_{\text{[NII]}}  T_{\text{H}\alpha}(\text{[NII]})}{ T_{\text{H}\alpha}(\text{H}\alpha)  T_{\text{[NII]}}(\text{[NII]}) -  T_{\text{H}\alpha}(\text{[NII]})T_{\text{[NII]}}(\text{H}\alpha)}\text{ ,}
\end{equation}

\noindent
with a similar equation for $f(\text{[NII]})$. Equivalent widths were then converted to rest frame using the redshift information. The distribution of H$\alpha$ fluxes can be seen in Fig. \ref{Fhalpha_AGN}. The median error was $\sim 12$\%, in agreement with the simulations performed by \cite{laralopez2011}, who obtained errors below 20\% for a FHWM of the OSIRIS TF of  12 \AA\ and a sampling of 6 \AA. However, our errors in the measurement of the [NII] line were much higher, with 60\% of the objects having errors above 50\%. This was to be expected since the [NII] line is usually fainter than the H$\alpha$ one, especially in a sample mainly composed of low-luminosity sources, as demonstrated in \cite{ramonperez2019}. As noted in the following section, SFGs constitute about two thirds of the 28 (from 46) H$\alpha$ sources with [NII] fluxes effectively measured after the BLAGN segregation.

\subsubsection{Discrimination between star-forming galaxies and AGNs}  
\label{DiscriminationSFGAGN}

One of the most used diagnostic diagrams to discriminate between SFG and AGN hosts is the Baldwin, Phillips \& Terlevich (BPT) diagram (\citeyear{Baldwin1981}), which uses the ratios of [OIII]/H$\beta$ and [NII]/H$\alpha$ emission lines. Other flux ratios, such as [SII]($\lambda$6716+$\lambda$6731)/H$\alpha$ or [OI]$\lambda$6300/H$\alpha$ are also useful for this purpose (\citealt{Baldwin1981}, \citealt{Veilleux1987}). Unfortunately, those lines are not always available, and their fluxes cannot always be measured. Simpler diagnostic diagrams are thus needed to separate distinct classes of objects. An alternative is to use the named EW$\alpha$n2 diagram, in which the [OIII]/H$\beta$ ratio of the BPT diagram is replaced with the EW of H$\alpha$ at rest-frame \citep{CidFernandes2010}. \par 

In the EW$\alpha$n2 diagram, star-forming and active galaxies occupy separate regions along the horizontal axis, while Seyferts and LINERs are differently distributed along the vertical axis. Several criteria can be used in order to select AGNs. \cite{Stasinska2006}, for instance, defines pure SFGs as those objects lying in the log [NII]/H$\alpha \leq -0.4$ region and AGNs as those with log [NII]/H$\alpha > -0.2$. In the intermediate region, hybrid objects having both star-formation and nuclear activity are located. A similar classification for AGNs is proposed by \cite{Ho1997}, while \cite{Kewley2001} are slightly more restrictive and consider pure AGNs to be those objects with log [NII]/H$\alpha > -0.1$. Moreover, a separation between LINERs and Seyferts can be traced at EW(H$\alpha$)=6 \AA\ (rest-frame), according to \cite{Kewley2006}.\par

\begin{figure*}[!htb]
\centering
\includegraphics[width=0.7\textwidth]{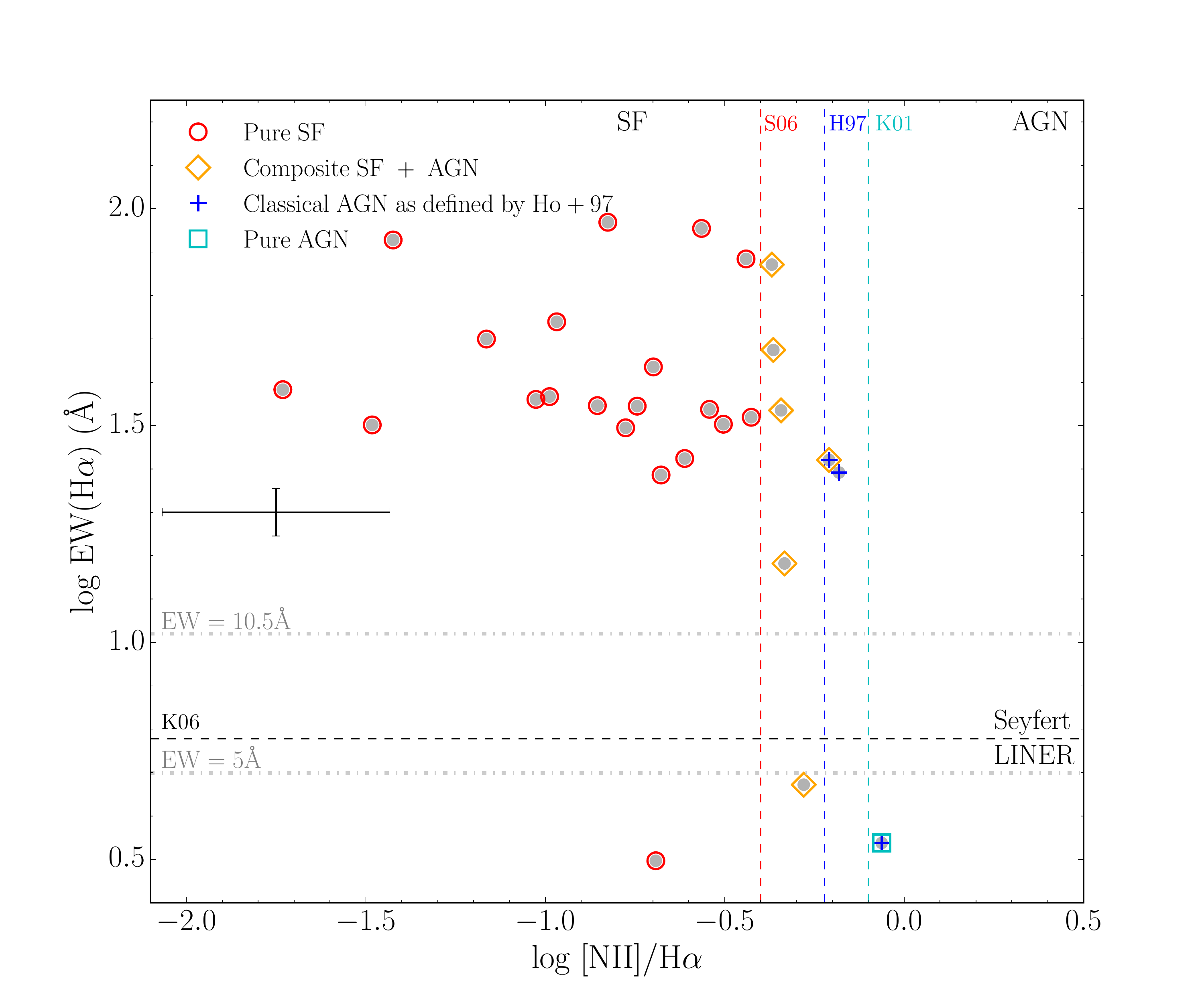}
\caption[EW$\alpha$n2 diagnostic diagram to distinguish star-forming galaxies from AGN]{EW$\alpha$n2 diagnostic diagram defined by \cite{CidFernandes2010} in order to distinguish SFGs from AGNs using emission lines in the optical. Pure SFGs (red circles) are separated from composite objects (SF+AGN, orange diamonds) according to \cite{Stasinska2006} (red dashed vertical line). Classical AGNs as defined by \cite{Ho1997} (blue dashed vertical line) are displayed with blue crosses. Pure AGNs according to \cite{Kewley2001} (cyan dashed vertical line) are represented in cyan squares. The black dashed horizontal line corresponds to the Seyfert/LINER separation criterion by \cite{Kewley2006}. The two grey dashed-dotted horizontal lines mark our minimum detected EW(H$\alpha$) with a probability threshold of $p\geq 0.95$ for objects with a PS continuum up to $\sim 10^{-18}$ erg s$^{-1}$ cm$^{-2}$ \AA$^{-1}$, and a EW(H$\alpha$) with $p\geq 0.50$ for objects with a PS continuum up to $\sim 10^{-19}$ erg s$^{-1}$ cm$^{-2}$ \AA$^{-1}$ \cite[see][]{ramonperez2019}. The error bars on the left-hand side of the plot represent the median of the relative errors of our H$\alpha$ sample in this space.}

\label{CidFernandes}
\end{figure*}

Figure \ref{CidFernandes} shows the EW$\alpha$n2 diagram for our sample of H$\alpha$ emitters. The different criteria previously described are shown. We have also traced our minimum detected EW(H$\alpha$) with a probability threshold of $p \geq 0.95$ and $p \geq 0.50$, according to the results of the simulations described in \cite{ramonperez2019}. In order to ensure the selection, objects with EWs below the $p \geq 0.95$ limit, including two possible LINERs, were discarded from the analysis. Accordingly, all the selected AGNs are presumably Seyfert galaxies with EW(H$\alpha \text{)}>0.6$. We selected all galaxies showing evidence of nuclear activity, either in composite (i.e. SF+AGN) galaxies or in pure active ones, following the criterion of \cite{Stasinska2006}. In this way, six H$\alpha$ emitters were selected as NLAGNs. However, due to the large uncertainties in the [NII] line flux measurements, the [NII]/H$\alpha$ ratios also have large errors (the median relative error in the sample was 32\%, see error bars in Fig. \ref{CidFernandes}) and therefore sources close to one side of the SF/AGN statistical frontier could belong to the another object type.\par 

We search for additional insights to reinforce the results of the EW$\alpha$n2 diagnostics. Only one of the H$\alpha$ emitters is included in the X-ray emitter subset (see Section \ref{demography_sec}). On the other hand, we have not found subtantial differences in optical/IR luminosity or MIR flux ratios between these subsamples of H$\alpha$ emitters with
an effectively measured [NII] line flux, as used in the following sections for the whole AGN population in OTELO presented here.\par

According to \cite{CidFernandes2011}, the BPT-based criterion of \cite{Stasinska2006} is more of a `pure-SF' demarcation line than a line used to divide SFGs from AGNs. In this sense, and taking into account the significance of the median error in the [NII]/H$\alpha$ ratio with respect to this boundary, more than half of the H$\alpha$ with [NII] line flux measured are bona-fide SFGs, and the selection of NLAGNs with this procedure should be taken with caution.

\subsection{X-ray selection} 
\label{XRayselection}

The strong X-ray emission, produced in the central regions of the accretion disc surrounding the black hole, is a good indicator of nuclear activity in galaxies. In particular, \cite{Maccacaro1988} showed the power of the X-ray-to-optical flux ratio (X/O) to distinguish AGNs from other X-ray-emitting sources. The OTELO catalogue has information in the soft 0.5-2 keV band from \cite{povic2009} and \cite{Laird2009}, and therefore we adopt the \cite{Szokoly2004} X/O definition:

\begin{equation}
\text{X/O} \equiv \log_{10}(f_\text{X}/f_\text{O}) \equiv \log_{10}(f_\text{X})+0.4\,\text{R}+5.71,
\label{XO}
\end{equation}

\noindent
where $f_\text{X}$ is the X-ray flux in the 0.5-2 keV band (erg s$^{-1}$ cm$^{-2}$) and R is the optical magnitude in Vega magnitudes. 

According to \cite{Stocke1991}, AGNs are typically located in the -1 < X/O < 1 range. At very high values of X/O we can find not only AGN types 1 and 2, but also clusters of galaxies at high redshift, extreme BL Lac objects, and cooling-flow galaxies. On the other hand, lower values of this ratio ($\text{X/O} < -1$) in extragalactic sources include normal and star-forming galaxies, as well as low-luminosity AGNs \cite[see][and references therein]{Alexander2001}, some of which would account for possible composite objects. For our purpose, objects with nuclear activity and those with star formation can be separated with the $\text{X/O}=-1$ limit.\par 

\begin{figure}[!htb]
\centering
\includegraphics[width=0.5\textwidth]{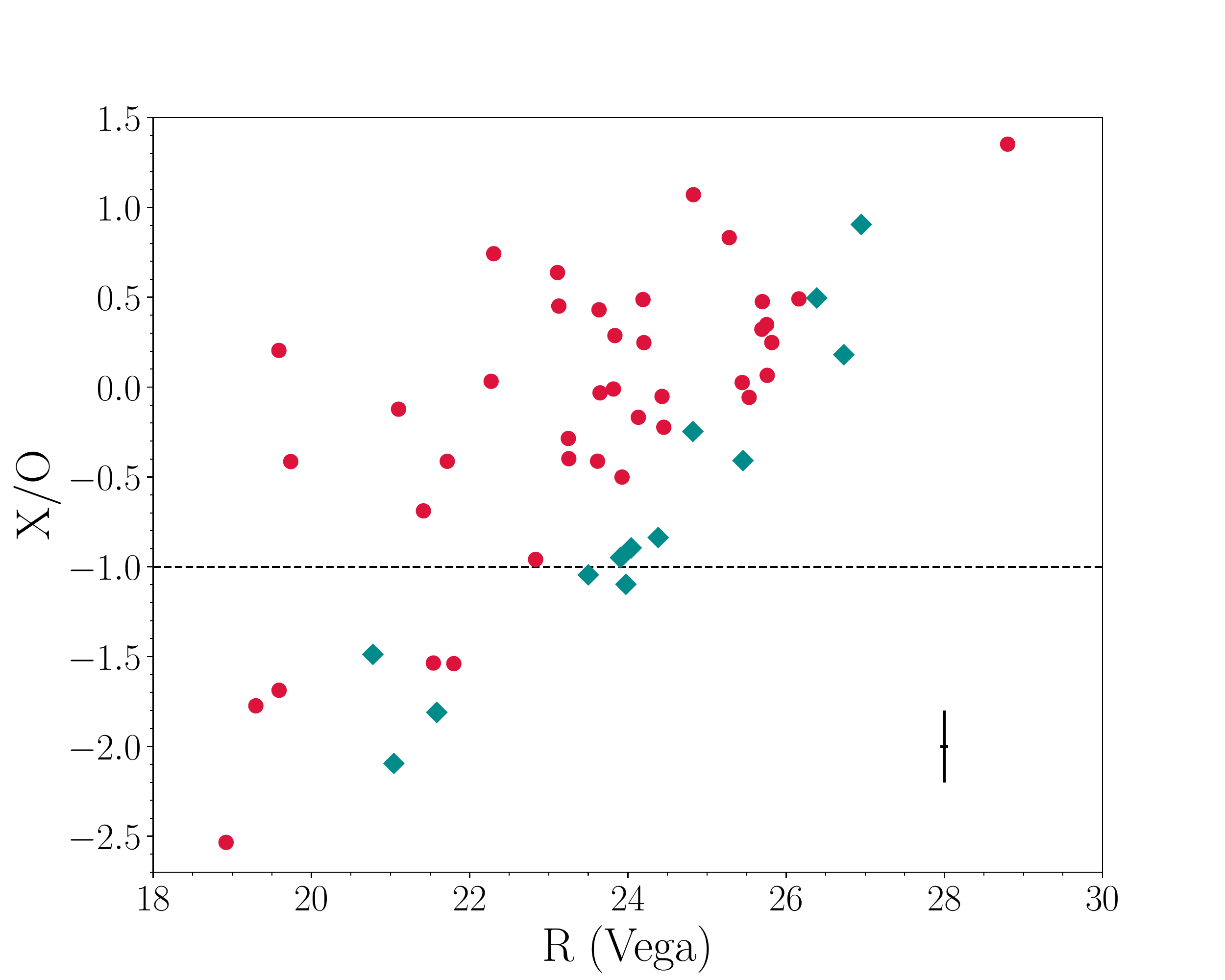}
\caption[X-ray-to-optical flux ratio as a function of the optical band]{X-ray-to-optical flux ratio (X/O) defined by Eq. \ref{XO}, as a function of the optical band (R in Vega magnitudes) of the X-ray emitting sources in the OTELO catalogue. Red dots correspond to the sources from the original catalogue of \cite{povic2009} with information in the soft band (0.5-2 keV), while the green diamonds are the ones from the catalogue of \cite{Laird2009}. The adopted criterion to select AGNs is $\text{X/O}>-1$ (dashed line). The bar represents the typical error propagated to the X/O ratio at this boundary for this sample. A total of 42 AGN counterparts was found regardless of the effects of this uncertainty estimation (see the text).
}
\label{XOR}
\end{figure}

In Fig. \ref{XOR} we plot the X/O ratio as a function of the optical magnitude, for the sources in the OTELO catalogue that show an X-ray emission and have information in the soft band (53 out of 56). We marked 42 sources with $\text{X/O}>-1$ as AGNs, but regarding the typical uncertainty in the calculation of this ratio, this number could vary by 5 to 7\% above or below this hard boundary, respectively, if the sources around it are different from low-luminosity AGNs \cite[see, for instance,][]{hornschemeier01} or composite objects.

\subsection{Mid-infrared selection}  
\label{MIRselection}

As mentioned in Section \ref{intro}, the SED of AGNs from UV to $\sim$5 $\mu$m is dominated by a power-law continuum, while SFGs tend to show a black-body spectrum that peaks at $\sim$ 1.6 $\mu$m as a signature of the underlying stellar populations. Thus, AGNs are redder than normal galaxies in the MIR, and IR colours can help to distinguish between different galaxy spectral types.\par

\begin{figure}[!ht]
\centering
\includegraphics[width=0.5\textwidth]{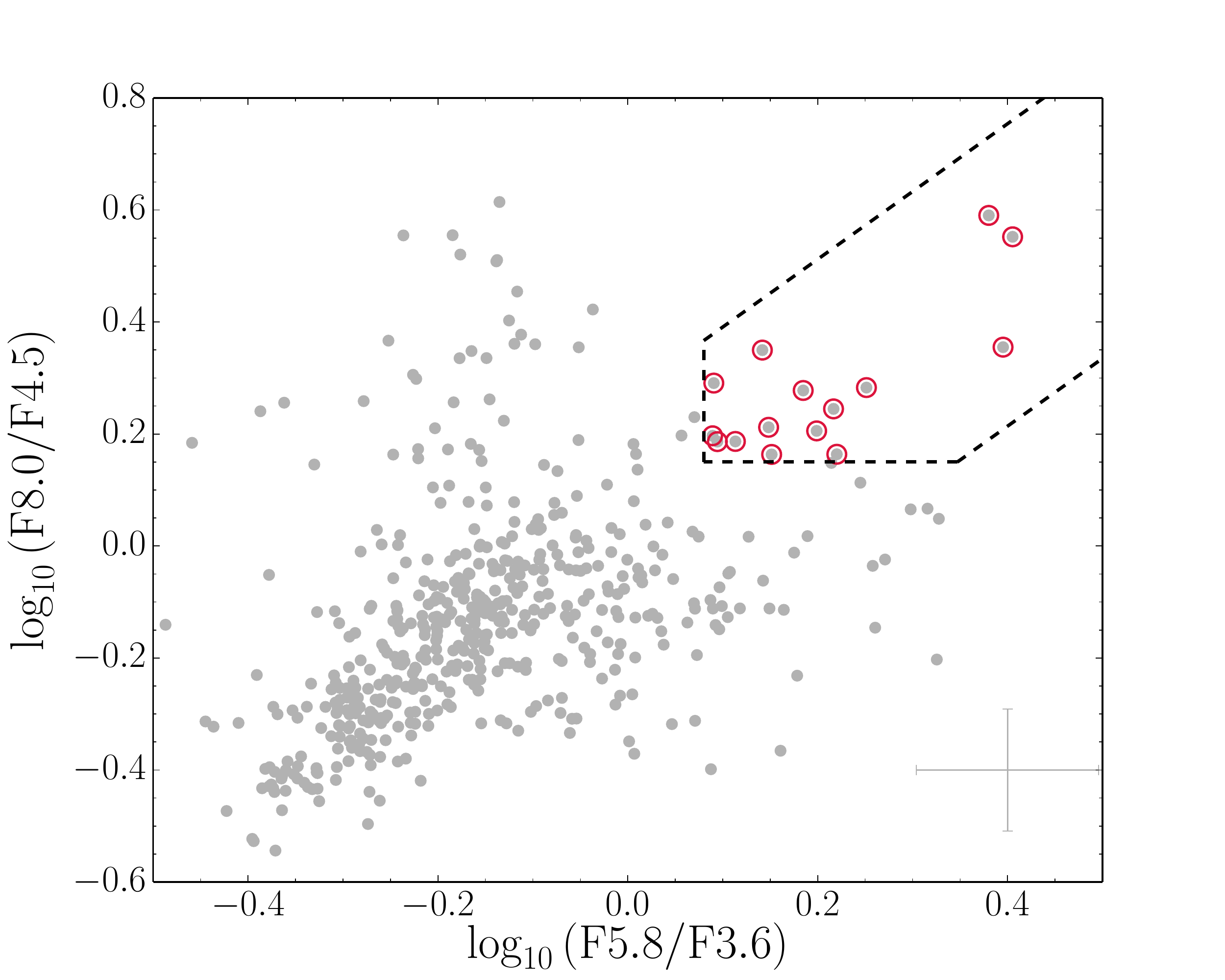}
\caption[Revised IRAC criteria from \cite{Donley2012} to separate AGN from star-forming galaxies]{Revised {\it Spitzer}/IRAC criteria from \cite{Donley2012} to separate AGNs from SFGs. The $x$ axis shows the ratio between the flux in the 5.8 $\mu$m band and the flux in the 3.6 $\mu$m band, while the $y$ axis depicts the ratio between the flux in the 8.0 $\mu$m band and the flux in the 4.5 $\mu$m band. Grey dots are all the sources in the OTELO catalogue with information in the four IRAC bands. Black dashed lines correspond to the limits set by \cite{Donley2012} to select AGNs. Red circled sources are the 15 sources selected as AGNs in this way.}
\label{donley}
\end{figure}

Diagnostic diagrams to discriminate AGNs from SFGs using IR colours are very common. One of the most remarkable is the empirical criterion proposed by \citealt{Stern2005}. Nevertheless, the authors claim that this method may omit AGNs at redshifts between $z\sim0.8$ and 2 and that the selection is contaminated by SFGs at high redshift. It is therefore not convenient in the case of OTELO, a survey that is not limited by redshift. Consequently, we decided to use the method of Donley et al. (2012) to select AGNs based on their MIR colours. This method makes use of the fluxes in the four {\it Spitzer}/IRAC bands (3.6, 4.5, 5.8 and 8.0 $\mu$m) and defines an empirical region where AGNs are found:

\begin{align}
\label{donley_eq}
\begin{split}
 x &\geq \text{ } 0.08 
\\
 y &\geq \text{ } 0.15 
\\
 y &\geq 1.21  \times  x -0.27
\\
 y &\leq 1.21  \times  x +0.27,
\end{split}
\end{align}

\noindent
where $x =\log_{10}(f_{5.8\,\mu \text{m}}/f_{3.6\,\mu \text{m}})$ and $y=\log_{10}(f_{8.0\,\mu \text{m}}/f_{4.5\,\mu \text{m}})$. As can be seen in Fig. \ref{donley}, 15 AGNs were found in this way.\par 

In addition, we use an alternative second MIR criterion to select AGNs in the field (the "KIM" criterion), based on the work by \cite{Messias2012}. This criterion not only takes the IRAC bands into account, but also the $K_{\rm s}$ and the 24 $\mu$m bands, which are also contained in the OTELO multiwavelength catalogue. The KIM criterion  defines the following region where AGNs are found (``IM'' criterion):

\begin{align}
\label{IM_eq}
\begin{split}
[8.0]-[24] &> -2.9 \times ([4.5] -[8.0])+2.8
\\
[8.0]-[24] &>0.5, 
\end{split}
\end{align}

\begin{figure}[!hbt]
\centering
\includegraphics[width=0.5\textwidth]{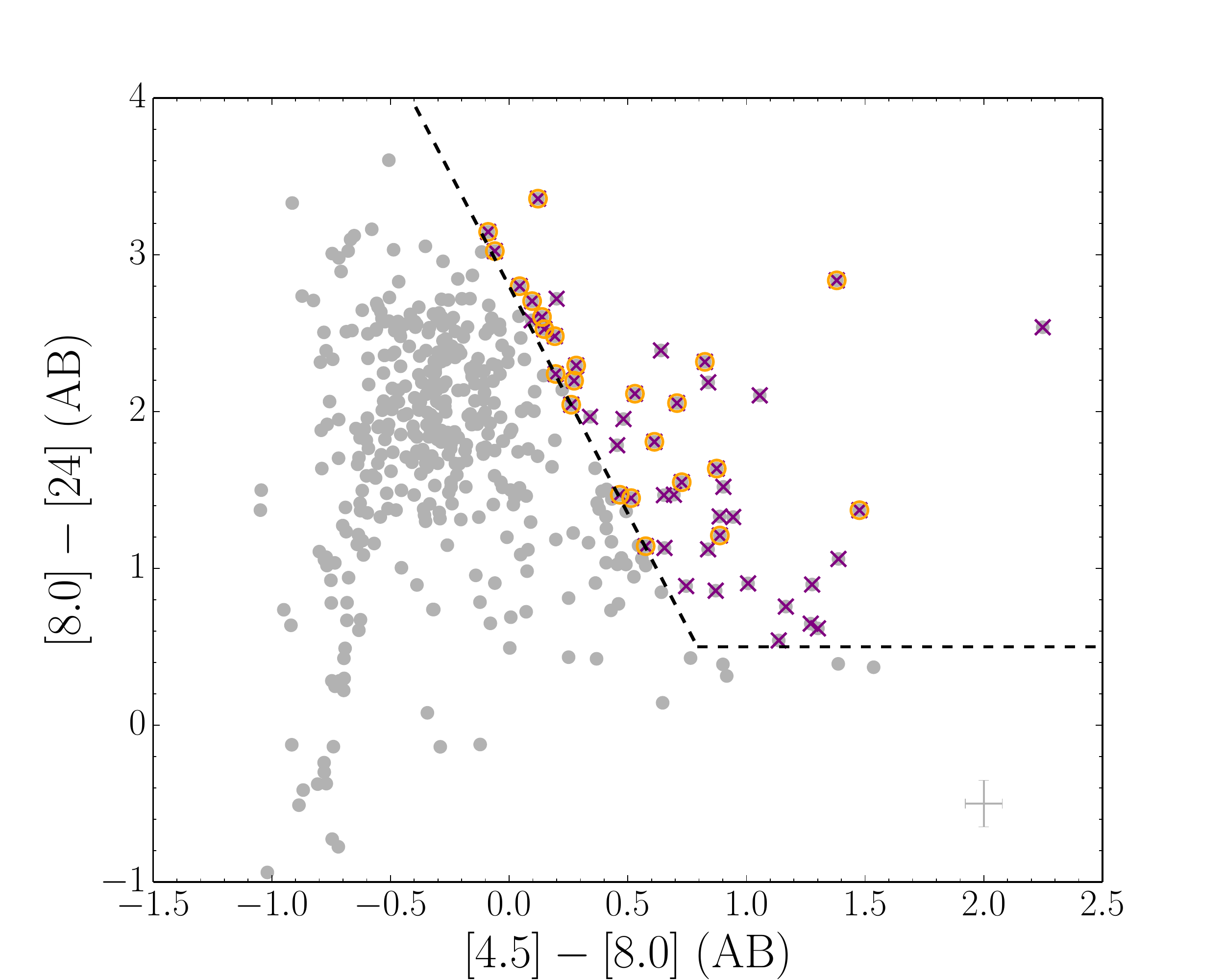}
\caption[IRAC+MIPS infrared criteria by \cite{Messias2012} to select AGN]{IRAC+MIPS (IM) IR criteria by \cite{Messias2012} to select AGNs. Grey dots are all the OTELO sources with information in the four IRAC bands as well as in the $K_{\rm s}$ and [24] $\mu$m bands. Black dashed lines are the limits of the IM criterion (see Eq. \ref{IM_eq}). Purple crosses represent the objects fulfilling the IM criterion. Orange circled sources represent the sources satisfying, in addition, that $K_{\rm s}-[4.5] > 0$, i.e., the KIM ($K_{\rm s}$+IRAC+MIPS) criterion. These are the objects selected as AGNs by the KIM criterion (24 sources).}
\label{KIM}
\end{figure}

\noindent
where [4.5], [8.0], and [24] represent AB magnitudes in the 4.5 and 8.0 $\mu$m IRAC bands and in the 24 $\mu$m {\it Spitzer}/MIPS band, respectively. In addition to that, sources have to fulfill a third condition: $K_{\rm s}-[4.5] >0$ (``K'' criterion). In this way, this method minimises contamination at low redshifts from normal galaxies while effectively separating AGNs from SFGs at high redshifts, and thus can be used at all ranges of $z$. In Fig. \ref{KIM}, we plot the $[8.0]-[24]$ versus $[4.5]-[8.0]$ colours, and have selected the sources fulfilling both the IM criterion and $K_{\rm s}-[4.5]  >0 $. Given that OTELO has no constraints on redshift, we used the KIM criterion and selected the latter (24 sources) as AGNs. From those, ten were selected by \cite{Donley2012} criteria and 14 were new. In total, 29 AGNs were selected using IR-based methods.\par \medskip

Finally, Table \ref{AGN_methods} summarises the different criteria used for the selection of AGNs, and the number of objects selected in each case. In total, 72 objects were classified as AGNs.

\begin{table}[ht]
\vspace*{2mm} 
\caption[Summary of AGN selection]{Summary of AGN selection. The first column indicates the number of objects in each group. The following columns specify the selection methods: X-rays, MIR, NLAGNs at $z=0.4$ or BLAGNs at $z=0.4$. The green checkmark means that an object at any redshift has been selected as an AGN by the corresponding method, while the red cross indicates that none of the objects have been selected by that method. Each row shows a subgroup of AGNs detected by one or more methods. The last row, in bold, indicates the total number of AGNs in each group.}       
\vspace*{-5mm}       
\label{AGN_methods}      
\centering     \small                                 
\begin{center}\begin{tabular}{c c c c c c}          
\hline   \\   [-0pt]                   
&  Number of & X-rays & MIR & NLAGNs & BLAGNs     \\  
&  objects &  &  &  ($z=0.4$) &  ($z=0.4$)     \\  
\hline      \\[1pt]                        
&    31 &  \textcolor{green}{\cmark} & \textcolor{red}{\xmark} & \textcolor{red}{\xmark} & \textcolor{red}{\xmark} \\[1pt] 
&    11 & \textcolor{green}{\cmark} & \textcolor{green}{\cmark} &\textcolor{red}{\xmark}  & \textcolor{red}{\xmark}      \\[1pt] 
&   18  & \textcolor{red}{\xmark}  & \textcolor{green}{\cmark} & \textcolor{red}{\xmark}  &  \textcolor{red}{\xmark}  \\[1pt] 
&   6  & \textcolor{red}{\xmark}  & \textcolor{red}{\xmark}  & \textcolor{green}{\cmark} & \textcolor{red}{\xmark}  \\[1pt] 
& 6  & \textcolor{red}{\xmark}  & \textcolor{red}{\xmark}  & \textcolor{red}{\xmark}  & \textcolor{green}{\cmark}  \\[6pt] 
\textbf{Total:} & \textbf{72} & \textbf{42} & \textbf{29} & \textbf{6} & \textbf{6} \\
\hline                                             
\end{tabular}\end{center}
\end{table}
\normalsize

\section{Analysis of AGNs} \label{analysisAGN}
Once the AGN population is selected from the OTELO survey, a first analysis is performed in order to gather information about its general characteristics, such as its demography and morphology. Also, the fraction of LIRGs and ULIRGs is studied. Finally, an inceptive analysis is performed for the subpopulation of AGNs at $z\sim 0.4$, which includes a study of the source environment. 

\subsection{Demography} 
\label{demography_sec}

\begin{figure}[!htb]
\centering
\includegraphics[width=0.5\textwidth]{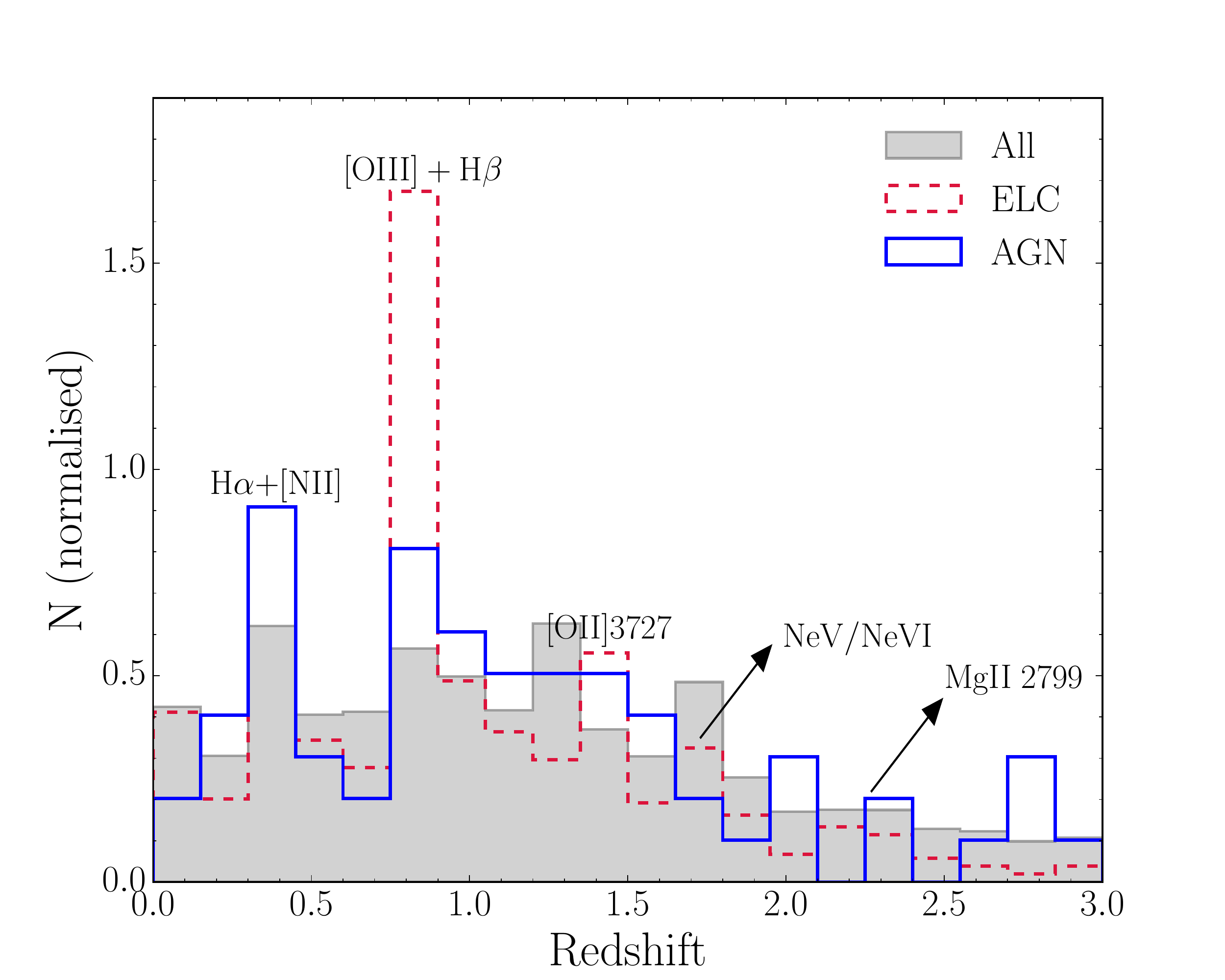}
\caption[Distribution of redshifts for all the objects in OTELO's catalogue, ELC and AGN]{Distribution of redshifts (obtained with LePhare as described in \citealt{OTELO1}) for the whole sample of OTELO (grey), the emitting-line candidates (ELC, shown in red; see text for details), and the AGN sample (shown in blue). Some of the most intense emission lines in this spectral interval are displayed.}
\label{redshifts_agn_elc}
\end{figure}

The AGN population found in OTELO with the methods described in the previous section comprises 72 objects and represents a very small fraction of the total number of objects in the catalogue (less than 1\%). Their distribution in redshift, as obtained with LePhare (see \citealt{OTELO1}), can be seen in Fig. \ref{redshifts_agn_elc}, together with that of the total and selected emission-line candidate (ELC) populations, which were selected using the methodology described in Sect. \ref{AGNat04}. It can be seen that there are more ELCs and AGNs at the redshifts corresponding to the more intense optical emission lines, which is a characteristic bias of emission-line surveys. In particular, the AGN population exhibits a peak at $z\sim0.4$, as this is the redshift at which the H$\alpha$ line appears in OTELO; we have focused on the search of those AGNs. This is not indicative of a redshift preference but rather a selection effect. It should also be noted that the proportion of AGNs over the total sample of objects is  higher at higher redshifts. This is expected since AGNs are very luminous objects and thus can be easily detected at higher redshifts. \par 

\begin{figure}[!hbt]
\centering
\includegraphics[width=0.5\textwidth]{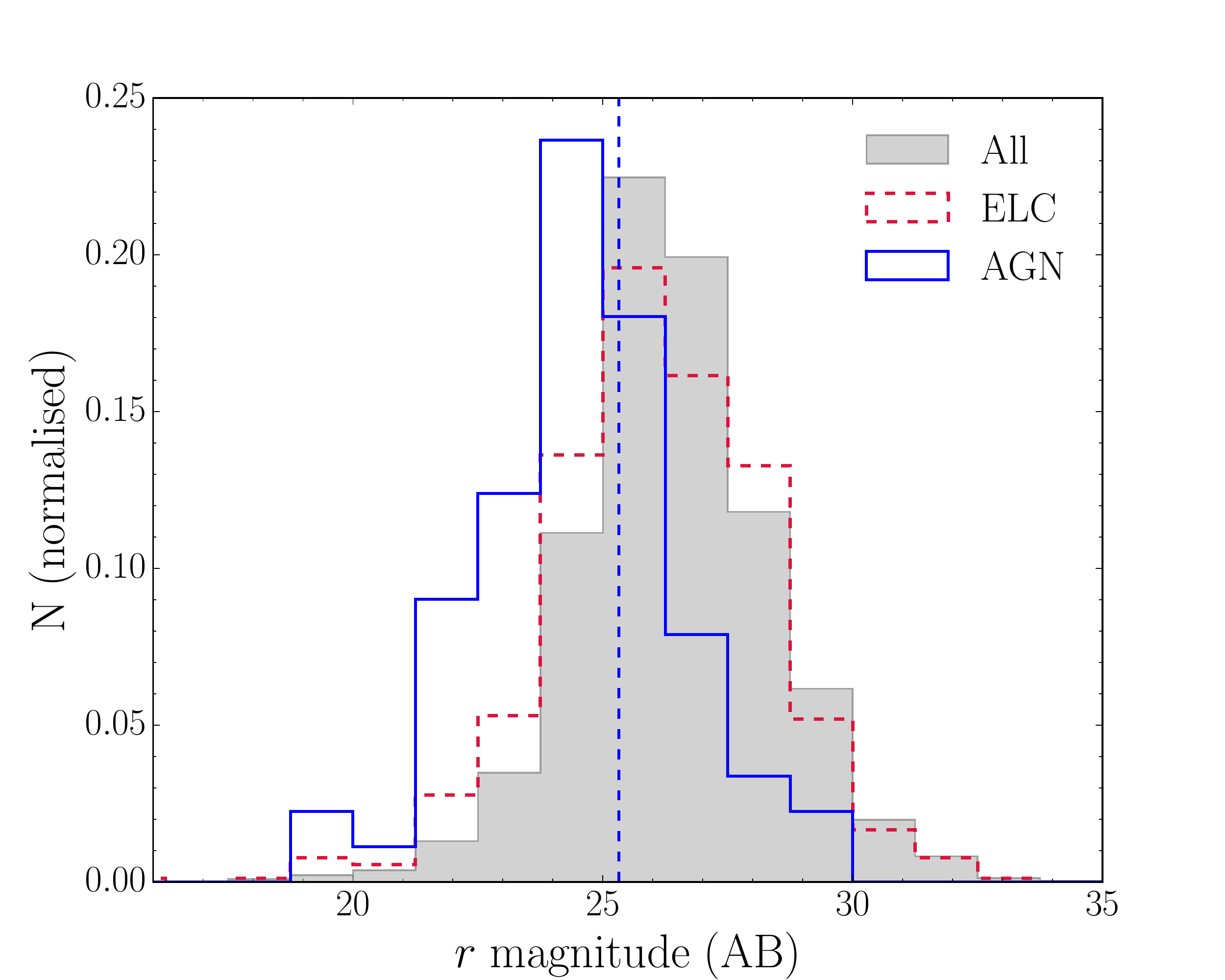}
\caption[Magnitude distribution in the $r$ band]{Normalised magnitude distribution in the $r$ band of all the objects in the OTELO catalogue (grey), the selected emitters (red dashed line), and the population of AGNs (blue solid line). The AGN distribution peaks at 24.5 mag, one magnitude and a half brighter that the two other distributions.}
\label{r_hist}
\end{figure}

Figure \ref{r_hist} represents the normalised distribution of magnitudes in the $r$ band, comparing the whole sample of OTELO with the selected emitters and the AGNs. As can be seen, the total and the ELC population show very similar distributions. Their median magnitudes are $26.4$ and $26.0\pm 2.2$, respectively. In the case of the AGN population, the distribution peaks at brighter magnitudes, the median being $24.5\pm 2.0$. This result was expected since the AGN phenomenon usually occurs in galaxies with higher luminosities than those with pure stellar formation \citep{osterbrock1991}. \par 
Below we summarise some of the characteristics of each AGN group according to their selection method.

\subsubsection*{X-ray-detected AGNs}

X-ray surveys are an efficient method to select AGNs, as can be deduced from Table \ref{AGN_methods}. A fraction of 43\% of our AGNs (31) were selected exclusively using the X/O ratio described in Sect. \ref{XRayselection}. In total, this method selected 58\% of the whole sample of AGNs (42). Moreover, of the 52 sources with X/O information, 81\% turned out to be AGNs, thus signaling that active galaxies could be responsible for the majority of the X-ray emission.\par 

X-ray-selected AGNs can be divided into two groups according to their level of obscuration caused by large columns of gas along the line of sight (N$_\text{H}>10^{22}$ cm$^{-2}$). In order to distinguish between unobscured and obscured X-rays AGNs, we used the hardness ratio, defined by \cite{povic2009} as follows: 

\begin{equation}
\text{HR}(\Delta E_1/ \Delta E_2) = \frac{\text{CR}(\Delta E_1)-\text{CR}(\Delta E_2)}{\text{CR}(\Delta E_1)+\text{CR}(\Delta E_2)},
\end{equation}

\noindent
where $\Delta E_1$ and $\Delta E_2$ are two different energy bands, in our case $\Delta E_1 = 2-4.5$ keV (hard2 band) and $\Delta E_2 = 0.5-2$ keV (soft band), and CR($\Delta E_{\rm n}$) is the count rate in the corresponding band. We used the criterion by \cite{DellaCeca2004}, who found that 90\% of their type 1 AGNs fell inside a narrow limit: $-0.75<\text{HR}<-0.35$, while type 2 sources occupied a broader range with $\text{HR}>-0.35$. \par

\begin{figure}[!htb]
\centering
\includegraphics[width=0.50\textwidth]{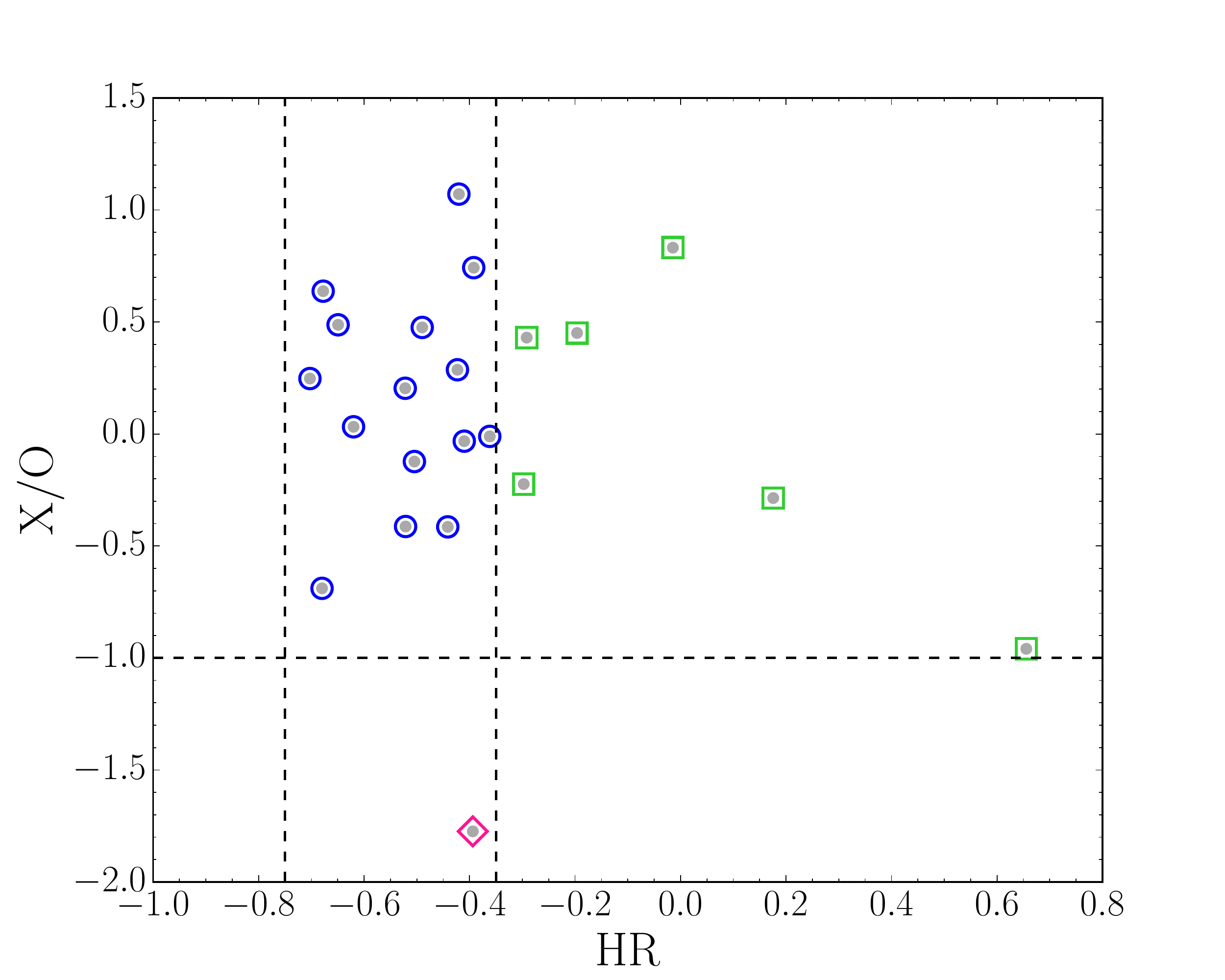}
\caption[X-ray-to-optical flux ratio as a function of the hardness  ratio]{X-ray-to-optical flux ratio (X/O) as a function of the hardness ratio, for the sources of \cite{povic2009} (grey dots) in the OTELO catalogue. The horizontal line corresponds to the limit $\text{X/O}=-1$, which separates AGNs ($\text{X/O}>-1$) from SFGs ($\text{X/O}<-1$). The two vertical lines correspond to the limits set by \cite{DellaCeca2004} which encloses type 1 AGNs ($-0.75 < \text{HR} < -0.35$), represented here by blue circles. Green squares are sources with  $\text{HR} > -0.35$, i.e. type 2 AGNs. The pink diamond represents a probable non-AGN source, which could be a coronal emitting star, a star-forming or early-type galaxy, or a heavily absorbed (Compton thick) AGN.}
\label{XOHR}
\end{figure}

In total, 21 out of our 42 X-ray-selected AGNs included the hardness ratio information in their catalogue listing. With the method described above, 15 sources were selected in the first category and 6 in the second. This represents a fraction of 71\% unobscured and 29\% obscured X-ray AGNs over the total subsample of those objects possessing information of their hardness ratio. This is in agreement with what was found by \cite{Marchesi2016}, where 69\% and 31\% of their whole sample of X-ray AGNs (both type 1 and type 2) were unobscured and obscured, respectively.

\subsubsection*{Active galactic nuclei selected based on mid-infrared}

The MIR selection methods described in section \ref{MIRselection} effectively selected 29 AGNs, 40\% of the sample. Eighteen of those objects, that is, a quarter of the AGN sample, were not selected by any other method. This implies that MIR selection is the second most effective method to select AGNs in this work. However, the fraction of IR AGNs over the total IR population is relatively small. In fact, barely 1\% of the objects for which information is available  in the four IRAC bands were selected as AGNs using the revised IRAC criteria from \cite{Donley2012} (see Fig. \ref{donley}). Similarly, of the objects for which information is available in the 4.5 and 8.0$\mu$m bands from IRAC and in the 24$\mu$m band from MIPS, only 3\% were classified as AGNs according to the KIM criteria of \cite{Messias2012}. The striking difference in the number of AGNs versus the total population of X-rays and MIR sources is also seen in the work of \cite{Cowley2016}, who performed a similar multiwavelength AGN selection (compare their Figs. 3 and 4). \par

As already mentioned, the great advantage of MIR selection is that it allows us to detect even those AGNs that are heavily obscured in X-rays. Consequently, by comparing the objects selected with MIR and X-rays methods, we can determine the fraction of obscured AGNs whose X-ray emission has been heavily absorbed by the surrounding interstellar gas or dust and re-emitted at IR wavelengths. In our case, 11 objects were selected both with MIR and X-rays methods while 18 were only selected with the former. This implies that 38\% of our IR AGNs are unobscured or moderately obscured and the rest (62\%) are heavily obscured. In their work, \cite{Mateos2012} selected AGNs with IR methods over the BUXS\footnote{Bright Ultra-hard XMM-Newton Survey.} field and found that 38.5\% had an X-ray counterpart, meaning they were not heavily obscured. This is in agreement with our findings.

\subsubsection*{AGNs at \textit{z} $\sim$ 0.4}

Our final sample of H$\alpha$ emitters at $z\sim 0.4$ was composed of 46 objects. From those, 12 were optically selected as AGNs (half of them being BLAGNs and half NLAGNs). From the rest of the AGN sample, only one object (X-ray-selected) fell at that redshift. In total, we have 13 AGNs at $z\sim 0.4$.\par 
In order to evaluate the proportion of line emitters and AGNs at that redshift, we first estimated the total number of objects found at $z\sim 0.4$ in OTELO. 
Considering an error of $\sim 0.2$ in the redshifts calculated with LePhare, as indicated in Section \ref{AGNat04}, we focused our search on the spectral window $0.37 < z < 0.42$, which covers the H$\alpha$ and [NII] lines in OTELO of $\pm 0.2$ in redshift. To these objects, we added the H$\alpha$ emitters that did not have a redshift in that interval but were classified as $z\sim0.4$ emitters by alternative methods. We avoided stars by discarding bright objects (with an AB magnitude in the deep image $<24$) with a stellarity index $>0.95$ from SExtractor. In total, the population of sources at $z\sim 0.4$ in OTELO was estimated to be  approximately 186 objects. This would imply that $\sim$25\% of the objects at $z\sim 0.4$ are line emitters, while $\sim$7\% are AGNs. However, due to the small sample size here, these values may not be statistically significant. Furthermore, the fraction of optically selected AGNs (NLAGNs or BLAGNs) over the sample of H$\alpha$ emitters is 26\% (see Table \ref{az04}).\par

\begin{table}[ht]
\vspace*{2mm} 
\caption[OTELO sources at $z\sim 0.4$ and fraction of emitters and AGN]{OTELO sources at $z\sim 0.4$ and fraction of emitters and AGNs. First column: Total number of OTELO sources at $z\sim 0.4$ (see text for details). Second column: Total number of H$\alpha$ selected emitters. Third and fourth columns:  Number of optically selected and X-rays-selected AGNs at that redshift. Fifth column: Total number of AGNs at that redshift. The second row shows the proportion of emitters and AGNs over the total sample of objects at that redshift. The third row indicates the proportion of optically selected AGNs (NLAGNs or BLAGNs) over the sample of emitters. Due to the small numbers that are being managed here, and the uncertainty in the estimation of the total number of sources at $z\sim 0.4$, these numbers, especially those in the second row, should be taken with caution.
}       
\vspace*{-5mm}       
\label{az04}      
\centering        \small                          
\begin{center}\begin{tabular}{c |  c |  c  c c}   
\hline
 & & & &  \\
\textbf{Objects at} & \textbf{H}$\alpha$  & & \textbf{AGNs} &     \\[3pt] 
\textbf{z}$\sim$ \textbf{0.4} & \textbf{Emitters}   & Optical & X-rays & Total      \\[3pt] \hline
 & & & &  \\
$\sim$186 & 46 & 12 & 1 & 13    \\[3pt]             
100\% & $\sim$25\% & & &  $\sim$7 \%      \\[3pt]             
 & 100\% & 26\% & &      \\[3pt]             
\hline                                             
\end{tabular}
\end{center}
\end{table}

\begin{figure}[!htb]
\centering
\includegraphics[width=0.50\textwidth]{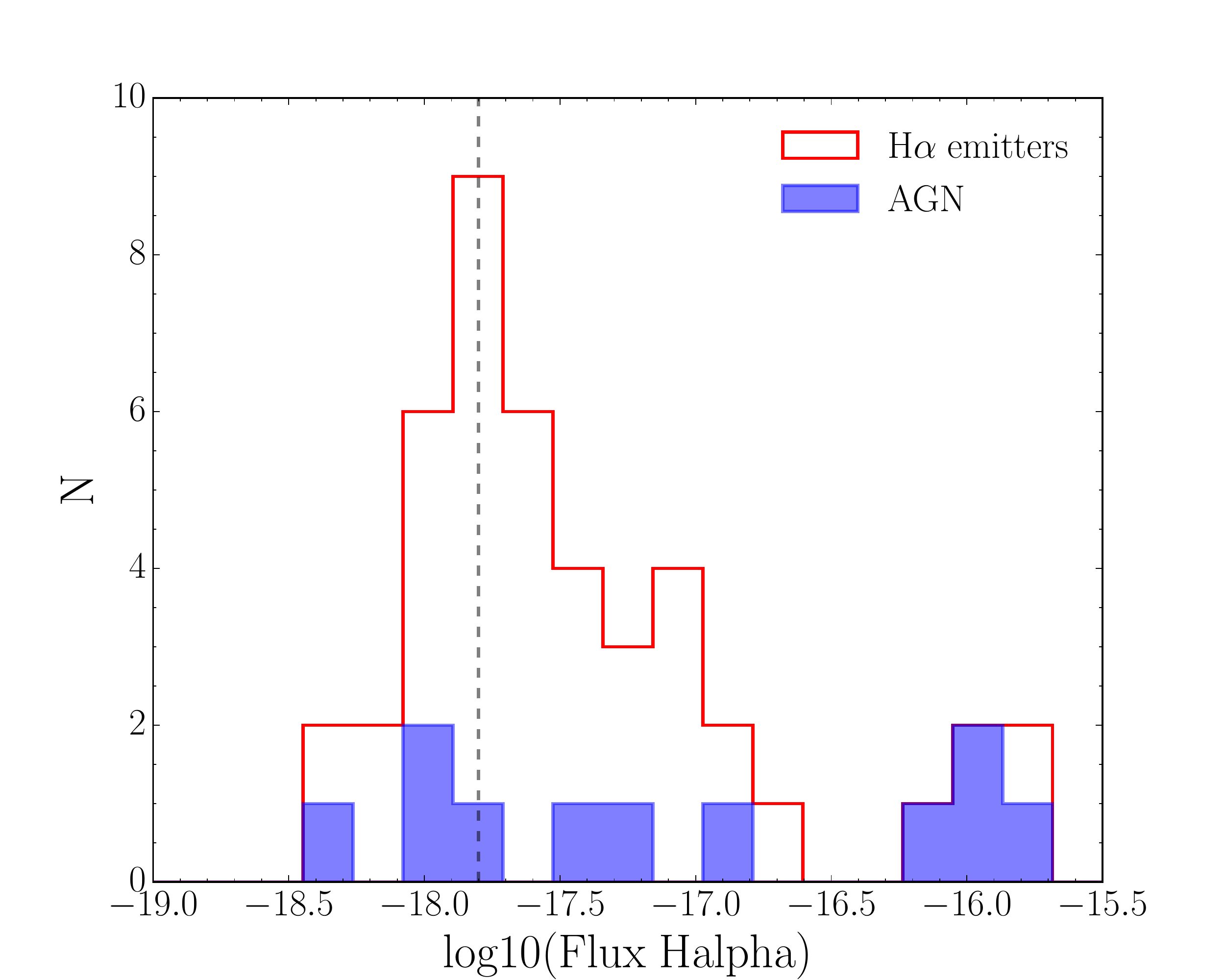}
\caption[Histogram of H$\alpha$ fluxes for the H$\alpha$ emitters and the AGN at $z\sim 0.4$]{Histogram of H$\alpha$ fluxes of the sources selected as H$\alpha$ emitters (red line) and the optically selected AGNs  at $z\sim  0.4$ (blue). The grey dashed line marks the peak of the distribution, corresponding to a flux of $\sim 1.6 \times 10^{-18}$ erg s$^{-1}$ cm$^{-2}$.
}
\label{Fhalpha_AGN}
\end{figure}

While the total number of sources at $z\sim 0.4$ may be subject to errors due to the uncertainty in our photo-$z$ calculations, especially for the faintest sources, the sample of H$\alpha$ emitters on the other hand was carefully inspected by different collaborators and therefore we fully rely on them. In their recent work from the HSC-SSP\footnote{Hyper Suprime-Cam (HSC) Subaru Strategic Program (SSP). See \href{http://hsc.mtk.nao.ac.jp/ssp/}{http://hsc.mtk.nao.ac.jp/ssp/}.}, \cite{Hayashi2018} found 14513 H$\alpha$ emitters in a total comoving volume of $9.77\times 10^5$ Mpc$^3$, using the NB921filter to select the objects. This volume is 508 times greater than that covered by the OTELO field in the redshift range $0.37 < z < 0.42$, which is $1924.31$ Mpc$^3$. According to these results, we would expect to find $\sim 29$ H$\alpha$ emitters in our field. This means we have found significantly (one third) more emitters in the OTELO survey. This difference may be attributed to the limiting line flux reached by the Subaru team ($1.5 \times 10^{-17}$ erg s$^{-1}$ cm$^{-2}$), which is higher than ours ($\sim 1.6 \times 10^{-18}$ erg s$^{-1}$ cm$^{-2}$, see Fig. \ref{Fhalpha_AGN}), and also to the fact that their detection method, based only on a colour-colour diagram, is less efficient. As a matter of fact, with this method they are only able to select objects with an observed $\text{EW}>25$ \AA\ for the NB921 filter, while our restrictions in EW go much lower.\par 

\cite{Sobral2013} also conducted a survey to find H$\alpha$ emitters at $z\sim 0.4$ using the NB921 filter and the same colour-colour diagram technique with an identical EW cut to that of the Subaru team. \cite{Sobral2013} found 1742 emitters over a cosmic volume of $8.8\times 10^4$ Mpc$^3$, 46 times bigger than our own. Translated to the OTELO volume, this would imply 38 emitters in our field, a value closer to what we find but still smaller. It is clear from these two comparisons that the potential of OTELO's pseudo-spectra to select emitters is noticeable.\par

As for the AGN fraction, we found that \cite{Sobral2013} and other authors estimated an AGN contribution to the H$\alpha$ population of $\sim$10\,--\,15\%, up to $z\sim 1$. This range is consistent with the results obtained from the analysis of emission-line galaxies at $z < 0.36$ from SDSS and GAMA surveys performed by \cite{laralopez2013}, who set this contribution to $\sim$11\% in each case. However, in a more recent work, \cite{Sobral2016} found that the AGN fraction strongly correlates with H$\alpha$ luminosity. While for low luminosities the previous estimation is acceptable, for higher luminosities the AGN fraction strongly increases. These latter authors estimated the AGN fraction to be 30\% and found that the most luminous H$\alpha$ emitters at any cosmic time are BLAGNs. In our case, we find a higher mean proportion of optically selected AGNs (26\% of the overall H$\alpha$ population, at any luminosity) although given our small numbers this is within the uncertainties (see Table \ref{az04}). On the other hand, our fraction of AGNs is almost 100\% at the highest luminosities, as shown in Fig. \ref{Fhalpha_AGN}, in agreement with  \cite{Sobral2016}. Moreover, our brightest AGNs are the broad-line ones, as also found by these latter-mentioned authors.

\subsection{Morphology} 
\label{section_morphology}

We studied the morphology of our AGNs using GALAPAGOS \citep{Barden2012}, a fully automated piece of software which combines the detection of objects with SExtractor and their light profile modelling with GALFIT \citep{Peng2002}. GALAPAGOS was run over the high-resolution images from the Hubble Space Telescope (F606W and F814W filters) corresponding to the OTELO field of view. The HST images cover the whole field of OTELO except for a $\sim$3.1 arcmin$^2$ region in the lower left-hand corner. The objects detected in this way were matched to the sources in the OTELO catalogue. GALFIT then obtained a light model with a Sérsic profile \citep{Sersic1963} for each of the detected components, starting by the brightest ones.

This model was then subtracted to the original image in order to obtain a residual image, showing possible hidden subcomponents of the object. Several examples of this procedure (original HST images used for the detection, Sérsic profile modelled by GALFIT and residual images) are shown in Appendix \ref{appendix_morphology}.\par 

Of the total sample of 72 AGNs, a GALFIT S\'ersic model with one or more components was obtained for 56 OTELO objects (and for their detected components in the high-resolution images). For the rest of them, either the source was so dim that it could not be fitted, or no HST image was available. We performed a visual classification of those objects by four collaborators based on the following parameters: 1) the appearance of the object in the HST images; 2) the GALFIT model and, in particular the value of the Sérsic index of the main component, $n$; 3) the existence (or not) of a residual after subtracting the model from the original image (revealing possible spiral arms, bars, and hidden structures); and 4) the relative colour of the source in a colour-colour diagram (such as $u-K_{\rm s}$ vs. $z-K_{\rm s}$, for instance). Based on that, each object was classed in one of the following categories: 1) point-like sources, 2) early-type sources (spheroidal objects, including ellipticals, E, and lenticulars, S0), 3) late-type sources (objects with disc, including spirals, S, and irregulars, Irr), and 4) unclassifiable objects. A summary of this classification is shown in Table \ref{Table_morph}.

\begin{table*}[ht]
\small
\vspace*{2mm} 
\caption[Morphological classification of OTELO's AGN]{Morphological classification of OTELO AGNs. The different types of AGNs, according to their selection method (X-rays, MIR, or BLAGNs/NLAGN at $z\sim 0.4$) are divided into four morphological categories: point-like sources, early-type (including ellipticals and lenticulars), late-type (including spirals and irregulars), and unclassifiable galaxies.}       
\vspace*{-5mm}       
\label{Table_morph}      
\centering                                
\begin{center}\begin{tabular}{c c c c c c c }   
\hline\\[-2pt]
 & X-rays  &  X-rays + MIR & MIR & BLAGNs & NLAGNs & \textbf{Total}     \\
  &   &  (unobscured)& (obscured) & at $z\sim 0.4$ & at $z\sim 0.4$  \\[3pt] \hline \\[-3pt]
  Point-like & 3 & 2 & 2 & 0 & 0 & \textbf{7} \\[3pt]             
  Early-type & 5 & 2 &  0 & 0 & 1 & \textbf{8} \\[3pt]             
 Late-type & 14 & 6 & 11 & 5 & 0 & \textbf{36}     \\[3pt]             
 Unclassifiable & 3 & 0 &  1 & 0 & 1 & \textbf{5}     \\[3pt]             
 \textbf{Total} & \textbf{25} & \textbf{10} & \textbf{14} & \textbf{5} & \textbf{2} &  \textbf{56}     \\[3pt]             
\hline                                             
\end{tabular}\end{center}
\end{table*}

The majority of our AGNs were classified as late-type objects (64.3\%). Of those, 12 were clearly spirals (21\%) such as the ones shown in Figures \ref{1873} and \ref{8762} of Appendix \ref{appendix_morphology}, and 9 were irregulars (16\%) (see, e.g. Fig. \ref{7800}). On the other hand, 14.3\% of our AGNs were classified as early-type objects, including two possible lenticulars such as the one shown in Fig. \ref{5495}. Finally, 12.5\% of the sample were point-like sources and thus possible QSOs (see, e.g. the objects in Fig. \ref{6173} and \ref{8351}) and 8.9\% could not be classified (such as the one in Fig. \ref{3854}). In addition to that, 9 of the AGNs were flagged as multiple objects (16\%), meaning that what was seen as a single object in OTELO was actually a system of multiple components as revealed by the HST images (see Fig. \ref{2146}), and 9 (16\%) were flagged as having possible interactions or mergers (see Fig. \ref{5662} or \ref{11168}). \par 

If we study the morphological classification of our AGNs according to the selection methods, we can see that 80\%\ of the BLAGNs at $z\sim 0.4$ are in spiral galaxies, while the only NLAGN that was classified at that same redshift is an early-type instead. On the other hand, X-ray- and MIR-selected AGNs seem to share similar morphologies, although the fraction of late-type galaxies among the X-ray selected AGNs (56\%) is smaller than among the MIR ones (79\%). This is in agreement with \cite{Griffith2010}, who found that their MIR-selected AGNs had a slightly higher incidence of being hosted by disc galaxies than the X-ray selected ones, although both had similar morphologies in general. These latter authors explained this according to the scenario proposed by \cite{Gabor2009}, where AGNs represent an intermediate stage between disc-dominated and bulge-dominated galaxies. \cite{Hickox2009} also found results in agreement with this evolutionary scenario, where galaxies evolve from blue, disc-dominated types with radiatively efficient AGNs (optical- and IR-bright) to red, bulge-dominated ones with less efficient AGNs (optically faint, radio-bright) following the growth of the stellar bulge and a quasar phase. In this context, AGNs tend to be selected in MIR when the accretion to the SMBH is more effective and the reprocessing of UV photons to MIR by the dust torus is significant, while they are better selected in X-rays when the accretion is less efficient. \par  

\(\)\begin{figure}[!hbt]
\centering
\includegraphics[width=0.5\textwidth]{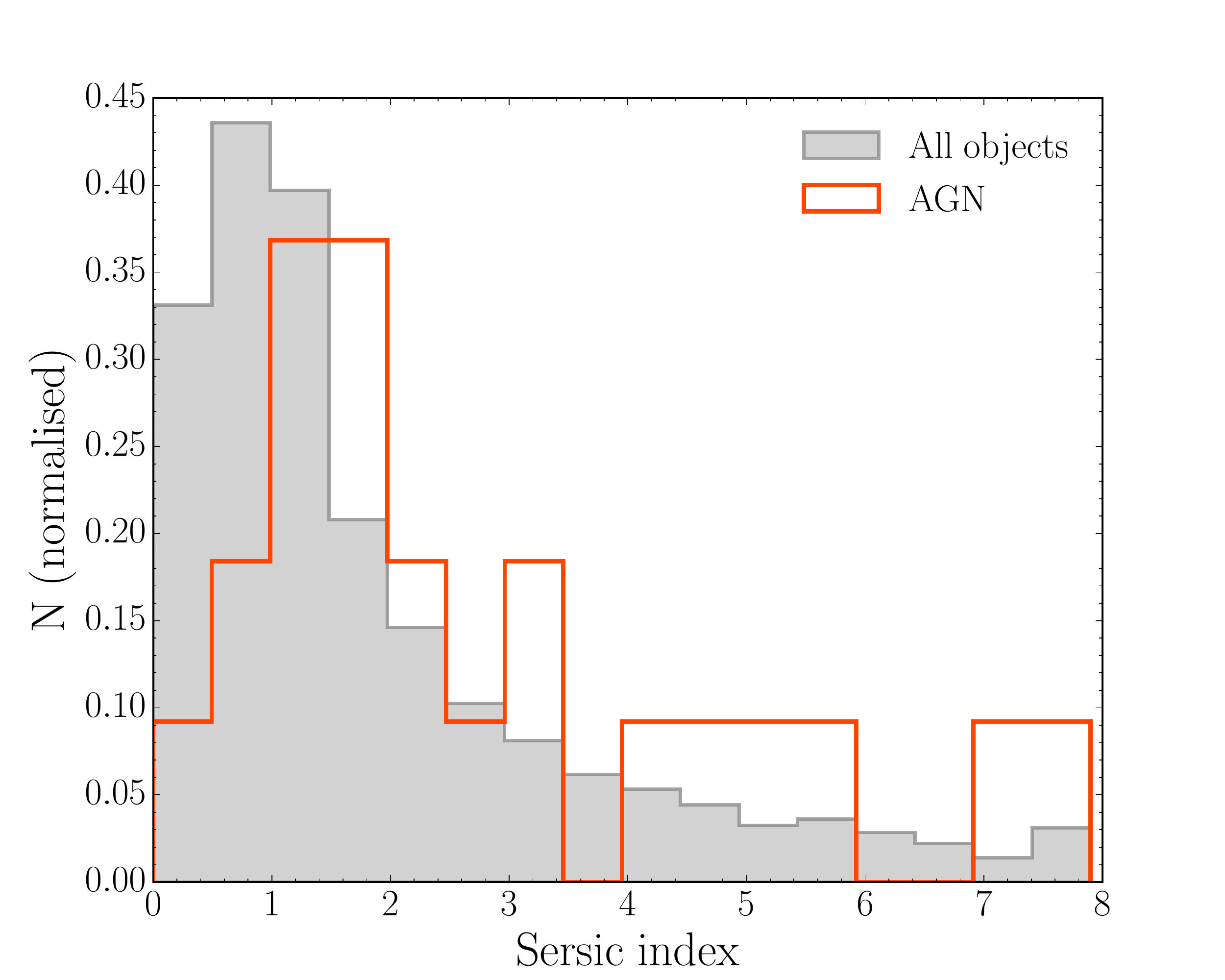}
\caption[Distribution of Sersic indices]{Sersic indices from the GALFIT models obtained for objects with unique components. The grey distribution represents the whole sample of OTELO sources that have been modelled as a single component (3286 objects). The orange distribution represents the AGN subpopulation.}
\label{hist_sersic}
\end{figure}

We also analysed the distribution of Sérsic indices obtained with the GALFIT models. According to the equation of a Sérsic profile, higher indices imply more concentrated objects with a steeper decrement in brightness. For that purpose, we used only the objects matched with the OTELO catalogue as individual sources and discarded the multiple ones, which would require a deeper and more detailed analysis. In Fig. \ref{hist_sersic} the distribution of S\'ersic indices is shown, both for the  whole sample of OTELO objects and for the AGN subpopulation. For the total population of OTELO objects, the distribution peaks at $n\sim 1$ and decreases steadily up to the maximum\textcolor{red}{\footnote{The minimum and maximum constraints set in GALAPAGOS/GALFIT were 0.2 and 8, respectively.}} value of $n=8$. The AGN distribution peaks at slightly higher indices and seems to have a higher proportion of concentrated objects. However, due to the small numbers being managed in the AGN sample, these variations may be attributed to statistical differences. As a matter of fact, \cite{Fan2014} performed a similar analysis over a sample of X-ray-selected AGNs at $z\sim 2$ and found  very similar AGN and non-AGN distributions to ours (see their Fig. 2). Their  numbers were also small (35 AGNs) and they concluded that there was no statistical difference between the AGNs and the control sample. In any case, the distribution of S\'ersic indices agrees with our previous findings about the predominance of late-type galaxies with a disc among the AGNs.

\subsection{Luminous and ultra-luminous infrared galaxies} 
\label{LIRGS_sec}

Luminous and ultra-luminous infrared galaxies are among the brightest galaxies in the Universe. As their names suggest, they emit most of their radiation in the IR, their luminosity in this range being superior to 10$^{11}$ and 10$^{12}$ L$_\odot$, respectively. The power source responsible for this emission is believed to be a starburst and/or an AGN. In order to look for these objects in OTELO, we first derived the IR luminosity (L$_{\text{IR}}$) of our sources. This luminosity is defined as the emission in the spectral range from 8 to 1000 $\mu$m. In this work, we took advantage of our multiwavelength catalogue and the fact that our sources may have not only MIPS photometry but also fluxes in the 100 and 160$\mu$m {\it Herschel}/PACS bands and in the 250, 350, and 500$\mu$m bands from {\it Herschel}/SPIRE.  We used the IR luminosity calculated by LePhare, which achieves this by integrating the emission in the range 8\,--\,1000 $\mu$m from the best FIR SED fitted to each galaxy.

\begin{figure}[!htb]
\centering
\includegraphics[width=0.5\textwidth]{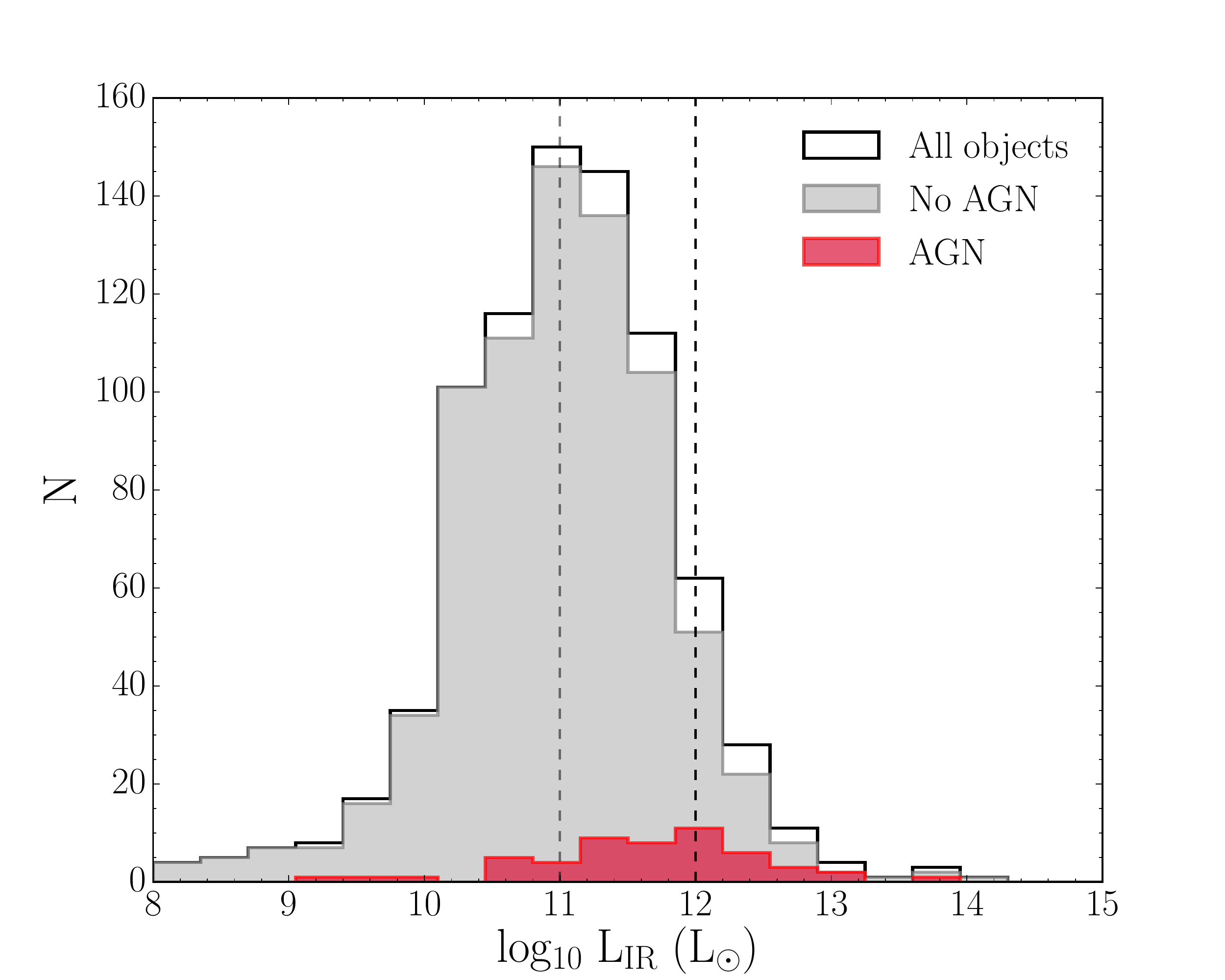}
\caption[Distribution of infrared luminosities]{Distribution of IR luminosities for the whole sample of OTELO sources for which 24$\mu$m photometry is available (black solid line), the objects not detected as AGNs (grey), and the selected AGNs (red). The grey and black dashed lines indicate the LIRGs and ULIRGs limit, respectively: 10$^{11}$ and 10$^{12}$ L$_{\odot}$.}
\label{Hist_Lir}
\end{figure}

\begin{figure}[!htb]
\centering
\includegraphics[width=0.5\textwidth]{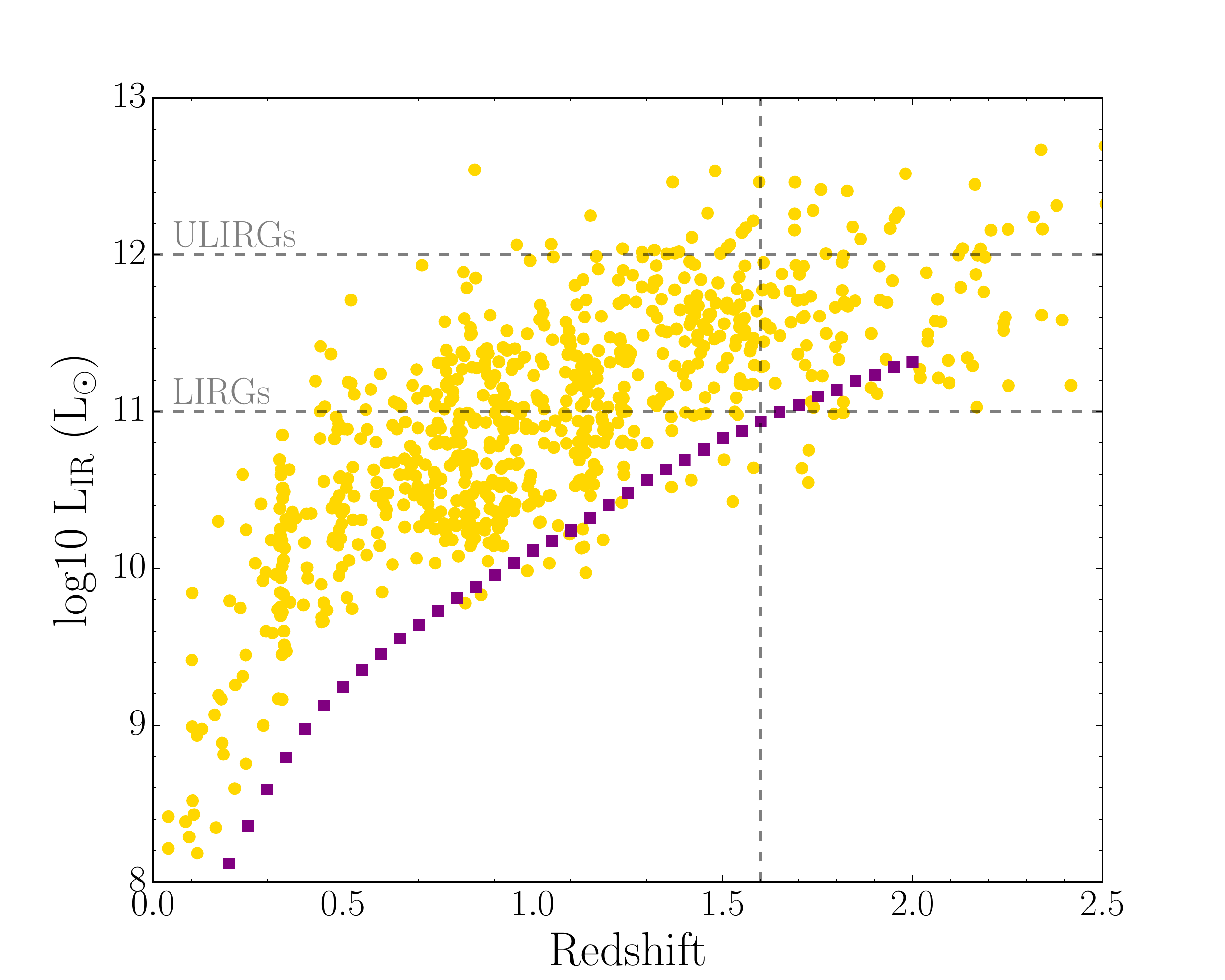}
\caption[Infrared luminosity in OTELO as a function of redshift]{Infrared luminosity in OTELO as a function of redshift. Yellow dots represent the IR luminosity of OTELO sources for which 24$\mu$m photometry is available. Purple squares represent the minimum detectable L$_{\text{IR}}$ given the sensitivity limits in our catalogue (see text for details). Horizontal lines indicate the LIRGs and ULIRGs limits. The vertical line shows the redshift up to which the sample of LIRGs is complete in our survey ($z=1.6$). }
\label{Min_Lir}
\end{figure}

The distributions of IR luminosities obtained in this way are shown in Fig. \ref{Hist_Lir}. As can be seen, at higher luminosities the proportion of AGNs contributing to the total L$_{\text{IR}}$ is greater. We can also deduce from the figure that the proportion of LIRGs and ULIRGs among AGNs is significant. However, in order to draw conclusions, we need to estimate the redshift up to which the sample of LIRGs and ULIRGs is complete. To do so, we used the minimum 24 $\mu$m MIPS flux detected in OTELO (21.95 AB) and a FIR SED template from \cite{CharyElbaz2001}. The template was redshifted from $z=0$ to 2.5 and re-escalated so that the minimum flux would correspond to the MIPS photometric point. Subsequently, the IR luminosity was obtained by integrating the SED flux from 8 to 1000 $\mu$m. The results are shown in Fig. \ref{Min_Lir}. As expected (see, e.g. Fig. 4 from \citealt{Elbaz2011}), the minimum detectable L$_{\text{IR}}$ increases with redshift. Given our sensitivity limits, we are able to detect all the LIRGs up to $z\sim1.6$, while the ULIRGs sample is complete at all redshifts.\par 

\begin{figure}[!hbt]
\centering
\includegraphics[width=0.5\textwidth]{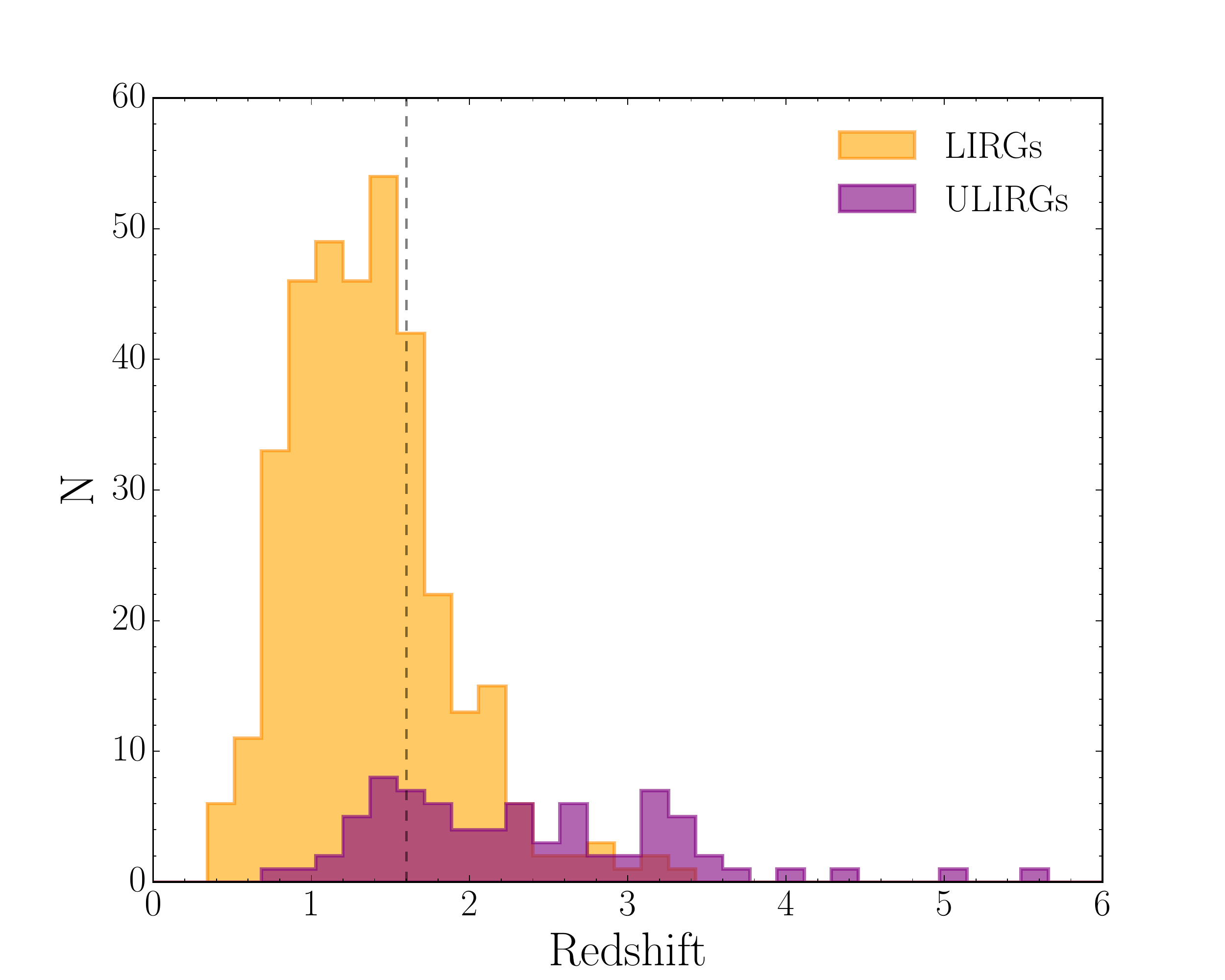}
\caption[Distribution of redshifts for LIRGs and ULIRGs]{Redshift distribution for LIRGs and ULIRGs (orange and purple, respectively). The grey dashed line indicates $z=1.6$, the redshift up to which the sample of LIRGs is complete.}
\label{Hist_Z_LIRGS_ULIRGS}
\end{figure}

The distribution of all the LIRGs and ULIRGs found in OTELO, as a function of redshift, is shown in Fig. \ref{Hist_Z_LIRGS_ULIRGS}. It can be seen that the number of LIRGs is much higher than that of  ULIRGs at low redshifts, and that it increases with $z$ (as also found by \citealt{Magnelli2013}, see its Fig. 12) up to $z\sim 1.5$, from where it starts to decrease as a consequence of our detection limits and in agreement with the previous estimation. At high redshifts, on the other hand, the number of ULIRGs prevails. Our distributions of LIRGs and ULIRGs over redshift are in agreement with what was found by \cite{Lin2016} and \cite{Malek2017}. \par 

Table \ref{Table_LIRGS_ULIRGS} details the number of ULIRGs found over the total sample of OTELO objects with L$_{\text{IR}}$, as well as among the AGN and non-AGN galaxies. We found that the fraction of LIRGs and ULIRGs is higher among the AGN population than among the rest of the galaxies, and that this difference is specially remarkable for the ULIRGs: 57\% (40\%) of AGNs (non-AGNs) are LIRGs, while 33\% (8\%) are ULIRGs. \par 

We also found that 8\% of LIRGs up to $z=1.6$ are active galaxies and that this number increases to 22\% for ULIRGs. This result is in agreement with \cite{Malek2017}, who found that ULIRGs are characterised by a higher fraction of AGNs than LIRGs. Among the 17 ULIRGs that are AGNs, all but one (21\%) were selected with MIR methods. This coincides with what was found by \cite{Veilleux1997}, who claimed that 25\,--\,30\% of their ULIRGs showed signs of activity in the optical or NIR.
\bigskip

\begin{table}[ht]
 \small 
\vspace*{2mm} 
\caption[Number of LIRGs and ULIRGs found in OTELO]{Number of LIRGs and ULIRGs found among the AGN and non-AGN populations, as well as the whole sample of OTELO sources with L$_{\text{IR}}$. The numbers in parentheses represent the fraction of each kind of object over the given population. (*): The LIRGs sample is studied only up to $z=1.6$ for completeness purposes (see text for details), while the ULIRGs sample covers all redshifts.}       
\vspace*{-5mm}       
\label{Table_LIRGS_ULIRGS}      
\centering                                
\begin{center}\begin{tabular}{c  r  r  r}   
\hline\\[-1pt]
 & LIRGs*  & ULIRGs & ULIRGs      \\[3pt] 
  & ($z<1.6$) & ($z<1.6$) & (all $z$) \\[3pt] \hline \\[-3pt]
AGNs & 21 (57\%) & 6 (16\%)  &  17 (33\%)   \\[3pt]             
non-AGNs & 243 (39\%) & 15 (2.4\%) &   60 (7.9\%)   \\[3pt]             
\textbf{Total} & 264 (40\%)  & 21 (3.2\%) & 77 (9.5\%)    \\[3pt]             
\hline                                             
\end{tabular}\end{center}
\end{table}

\subsection{AGN environment at z$\sim$0.4} 
\label{density_maps_sec}

Following a general analysis of the AGNs found in OTELO, we now focus on the population of objects at $z\sim 0.4$. We studied the environments of AGNs at this redshift in an attempt to determine whether they tend to be in high- or low-density environments. For details of the luminosity function (LF) of all sources of this population in the OTELO field, as well as the contribution of the AGN hosts to this LF, we refer the reader to \cite{ramonperez2019}.

\begin{figure}[!htb]
\centering
\includegraphics[width=0.5\textwidth]{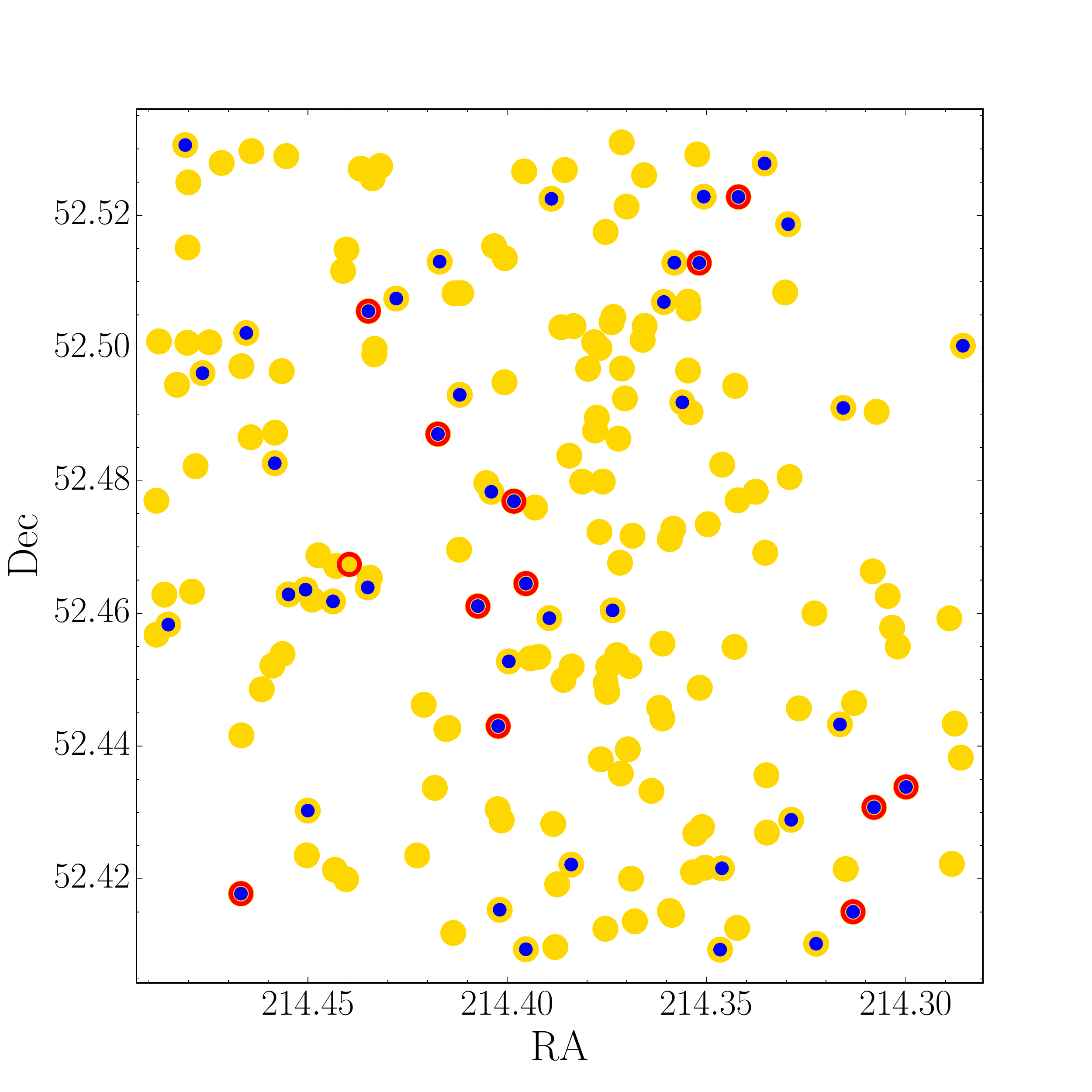}
\caption[Distribution of sources at $z\sim0.4$ in OTELO]{Distribution of sources at $z\sim0.4$ in OTELO. Filled yellow circles represent all the sources at that redshift. Filled blue circles are the H$\alpha$ emitters, while red circles are the selected AGNs.}
\label{Density_z040}
\end{figure}

The role of environment in AGN triggering is one of the most important open questions in the field. Both the large-scale and the local environments seem to be decisive in the evolution and properties of galaxies. In this section, we attempt to study the environments of AGNs at $z\sim0.4$. The spatial distribution  of all the sources at this redshift is shown in Fig. \ref{Density_z040}, along with that of the H$\alpha$ emitters and AGNs.\par 

 We studied the surface density by means of the projected fifth-nearest-neighbour distance ($D_{5}$) of each source. This method provides the most accurate estimate of local galaxy density when compared to other techniques such as the use of counts in a fixed aperture or the Voronoi volume, according to \cite{Cooper2005}. We used the edge correction by \cite{Kovac2010} to avoid obtaining artificially lower densities in the regions close to the edges of the field. The surface density is calculated as

\begin{equation}
\Sigma_5 = \frac{5}{\pi\,D^2_{5}}.
\end{equation}

\begin{figure}[!htb]
\centering
\includegraphics[width=0.5\textwidth]{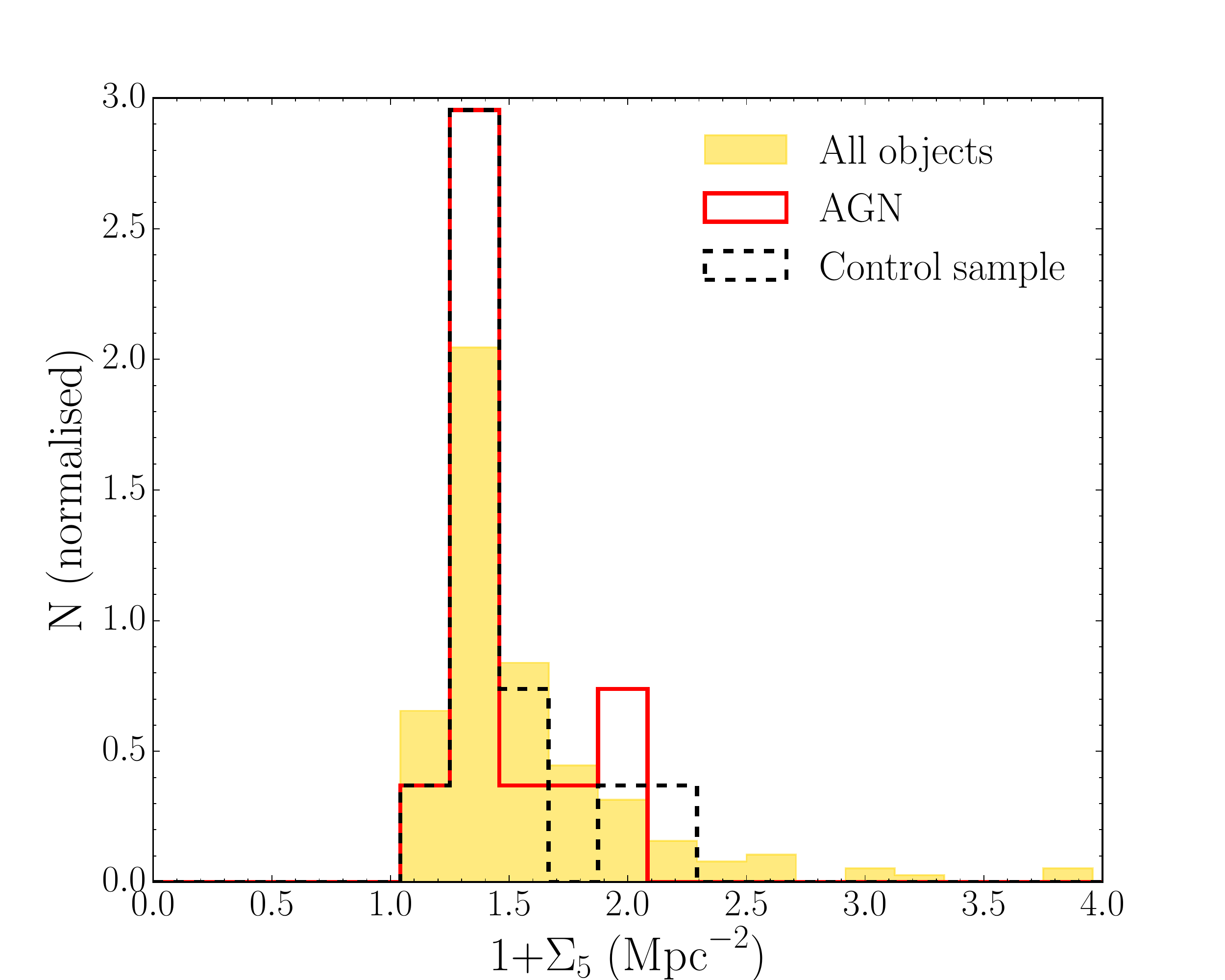}
\caption[Distribution of the surface density parameter ($\Sigma_5$) I]{Distribution of the surface density parameter ($\Sigma_5$) for the whole sample of OTELO objects at $z\sim0.4$ (yellow), the AGNs (red) and a random control sample of non-AGNs (black).}
\label{sigma_hist_new_5Mpc}
\end{figure}

We compared the distributions of the surface-density parameter for the sample of AGNs at $z\sim0.4$ (13 objects) and for a control sample of 13 non-AGNs randomly chosen at the same redshift. As can be seen in Fig. \ref{sigma_hist_new_5Mpc}, galaxies tend to concentrate in low-density environments. The AGN and non-AGN distributions are very similar, and seem to be in agreement with the density distribution of the overall population. We performed the analysis for different control samples and obtained analogous results every time.

\begin{figure}[!htb]
\centering
\includegraphics[width=0.5\textwidth]{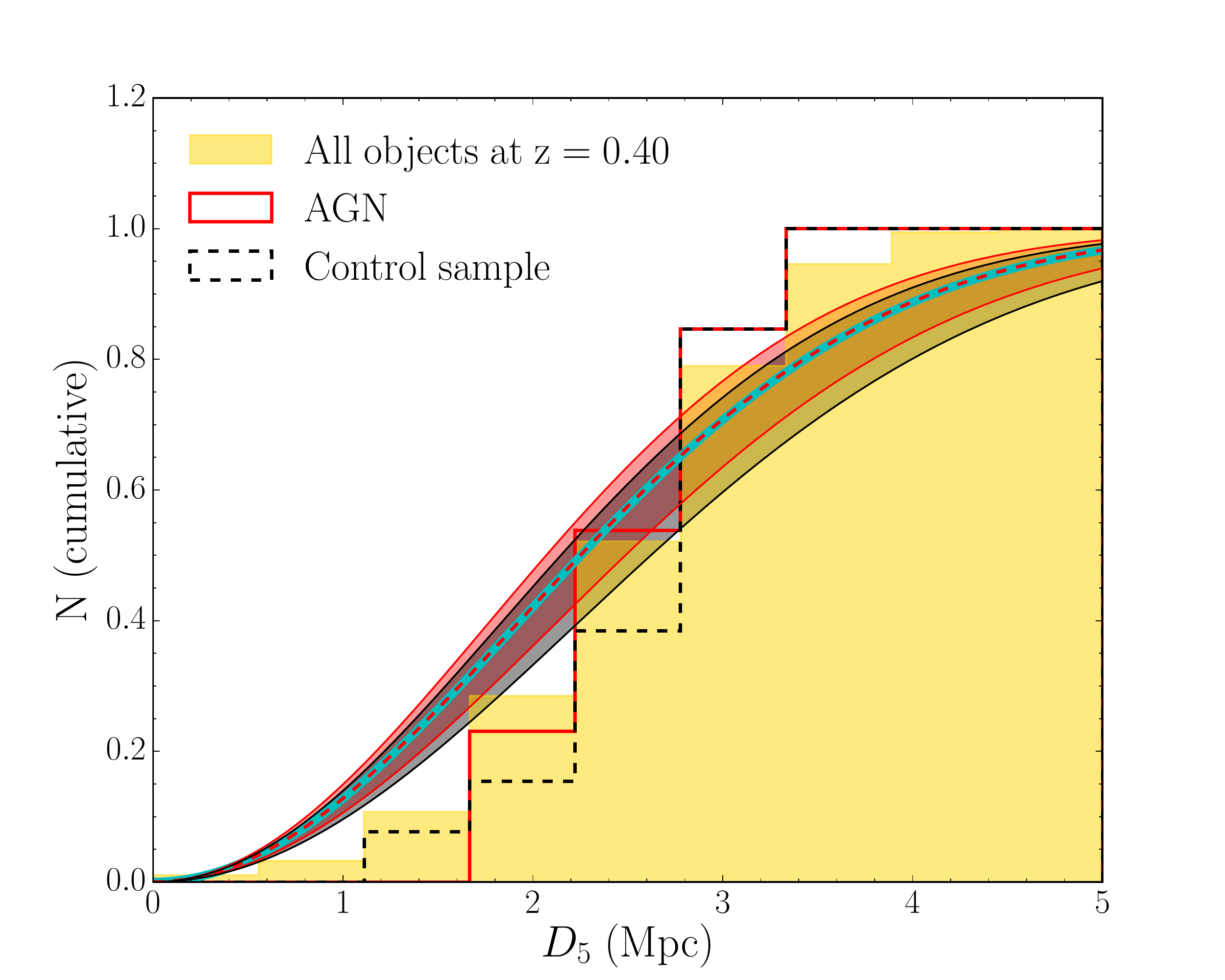}
\caption[Distributions of $D_5$ for the total sample of OTELO's objects, the AGN and the control sample (non-AGN)]{Cumulative and normalised distributions of $D_5$ for the total sample of OTELO objects, the AGNs, and the control sample (non-AGNs). The yellow distribution is the distribution of all the objects at $z\sim0.4$ and the cyan solid line is its Poissonian fit. The step histograms in red and black represent the distributions of $D_5$ for the AGN and non-AGN samples, respectively, while the corresponding dashed lines are their Poissonian fits. The filled bands show the propagation of statistical errors in the Poissonian fit.}
\label{poiss_ajuste_new_5Mpc}
\end{figure}

To further analyse the environmental differences between the AGNs and the control sample, we looked at the distance parameter, $D_5$, in each sample. The distance from a randomly chosen source to the fifth-nearest other object follows a homogenous Poisson process \citep{Martinez2002}  whose probability distribution function in 2D is given by
 
 \begin{equation}
G(D_{5}) = 1-\exp\,(-\rho\pi\, D_{5}^2),
\label{Gfunc}
\end{equation}

\noindent
where $\rho$ is the intensity of the process, or the expected  value of $D_5$. We fitted the cumulative distributions of $D_5$ in each sample to this function, assuming statistical errors of Poissonian nature, and compared it to the distribution of the overall population, composed of all the objects at $z\sim0.4$. The results are shown in Fig. \ref{poiss_ajuste_new_5Mpc}. It is clear from this figure that the AGN population and the control group do not statistically differ. This indicates that AGNs in OTELO field at $z\sim0.4$ are found in identical density environments to non-AGNs.\par 

\section{Conclusions}
\label{conclusions}

This work has focused on the identification and characterisation of AGNs in the OTELO survey, an emission-line object survey that uses the red tunable filter of the OSIRIS instrument at the GTC \citep{OTELO1}. We have obtained a sample of 72 AGNs in the OTELO field, selected using four different methods: an optical diagnostic diagram based on the H$\alpha$+[NII] fluxes of the emitters at $z\sim0.4$, a selection of BLAGNs at the same redshift based on the width of the emission line as seen in the pseudo-spectra, an X-ray selection, and two diagnostic diagrams in the MIR. \par 

The main goal of this work is to identify the main properties of the selected AGN population in OTELO. A detailed study on their demography, morphology, IR luminosity, and environment has been conducted. The main results from this analysis are summarised in the following paragraphs.\par

Regarding the AGN selection, X-ray emission has demonstrated to be the most efficient method, as similarly found by \citet{Mushotzky2004}. This method  selected 58\% of the whole sample of AGNs in OTELO. Around one third of these X-ray-selected AGNs are obscured. Mid-infrared diagnostic diagrams are also very effective, selecting 40\% of the OTELO AGN sample. In this case, the proportion of obscured and unobscured AGNs is 62\% and 38\%, respectively. This is roughly the opposite to what is found by selecting with X-rays. Both results are in agreement with previous works, such as \citet{Mateos2012} and \citet{Marchesi2016}.\par 

In the optical, at $z\sim0.4$ we found up to 13 AGNs, which represent 26\% of the H$\alpha$ emitters at that redshift. Compared to \citet{Sobral2016}, we found a higher mean proportion of optically selected AGNs. Following morphological criteria, the majority of our total sample of AGNs (64.3\%) were classified as late-type galaxies; this percentage includes a 16\% fraction of irregulars. A 14.3\% fraction were classified as early-type and 12.5\% as point-like sources, while 8.9\% could not be classified. Moreover, a 16\% fraction of the total sample show signs of interactions or mergers. Furthermore, most of the BLAGNs at $z\sim0.4$ are spiral galaxies (4 out of 5). Finally, the fraction of late-type galaxies among the X-ray-selected AGNs (56\%) is smaller than among the MIR-selected ones (79\%). This result is in agreement with \citet{Griffith2010} and the evolution scenario proposed by \citet{Gabor2009}, where AGNs represent an intermediate stage between disc-dominated and bulge-dominated galaxies.\par 

According to their IR luminosity, we are able to recover all the LIRGs in the OTELO field up to $z\sim1.6$. As expected, the fraction of LIRGs and ULIRGs is higher among the AGN population than among the rest of the galaxies, and this difference is particularly remarkable for the ultra-luminous type. However, the population of ULIRGs contains a higher fraction of AGNs than that of the LIRGs. Similar results were found by other recent works such as \citet{Lin2016} and \citet{Malek2017}. Active galactic nuclei in the OTELO field at $z\sim0.4$ are found in identical environments as non-AGNs (in agreement with \citealt{Virani2000} or \citealt{Waskett2005}) but the subpopulation of AGNs at $z\sim0.4$ is the most clustered group when compared to passive galaxies, SFGs, and H$\alpha$ emitters, something also observed by \citet{Manzer2014}.

\begin{acknowledgements} This  work  was  supported  by  the  Spanish  Ministry  of  Economy  and  Competitiveness  (MINECO) under  the  grants  
AYA2013\,-\,46724\,-\,P,
AYA2013\,-\,42227\,-\,P,
AYA2014\,-\,58861\,-\,C3\,-\,1\,-\,P,
AYA2014\,-\,58861\,-\,C3\,-\,2\,-\,P,
AYA2014\,-\,58861\,-\,C3\,-\,3\,-\,P,
AYA2016\,-\,75808\,-\,R,
AYA2016\,-\,75931\,-\,C2\,-\,1\,-\,P,
AYA2016\,-\,75931\,-\,C2\,-\,2\,-\,P,
AYA2016\,-\,76682C3\,-\,1\,-\,P,
AYA2017\,-\,88007\,-\,C3\,-\,1\,-\,P and
AYA2017\,-\,88007\,-\,C3\,-\,2\,-\,P.

The authors thank the anonymous referee for her/his feedback and suggestions.
Based on observations made with the Gran Telescopio Canarias (GTC), installed in the Spanish Observatorio del Roque de los Muchachos of the Instituto de Astrof\'isica de Canarias, in the island of La Palma. MP acknowledges financial supports from the Ethiopian Space Science and Technology Institute (ESSTI) under the Ethiopian Ministry of Innovation and Technology (MInT), and from the State Agency for Research of the Spanish MCIU through the "Center of Excellence Severo Ochoa" award for the Instituto de Astrof\'isica de Andaluc\'ia (SEV-2017-0709). EJA acknowledges financial support from the State Agency for Research of the Spanish MCIU through the "Center of Excellence Severo Ochoa" award for the Instituto de Astrof\'isica de Andaluc\'ia (SEV-2017-0709). 

\end{acknowledgements}

\bibliographystyle{aa}
\bibliography{otelo_agn_v3}

\newpage

\begin{appendix} 

\onecolumn

\section{Broad-line AGNs at $z\sim0.4$ in OTELO}\label{appendix_BLAGN}
\vspace{20pt}

\begin{minipage}[c]{\linewidth}
\centering
\includegraphics[width=0.95\textwidth]{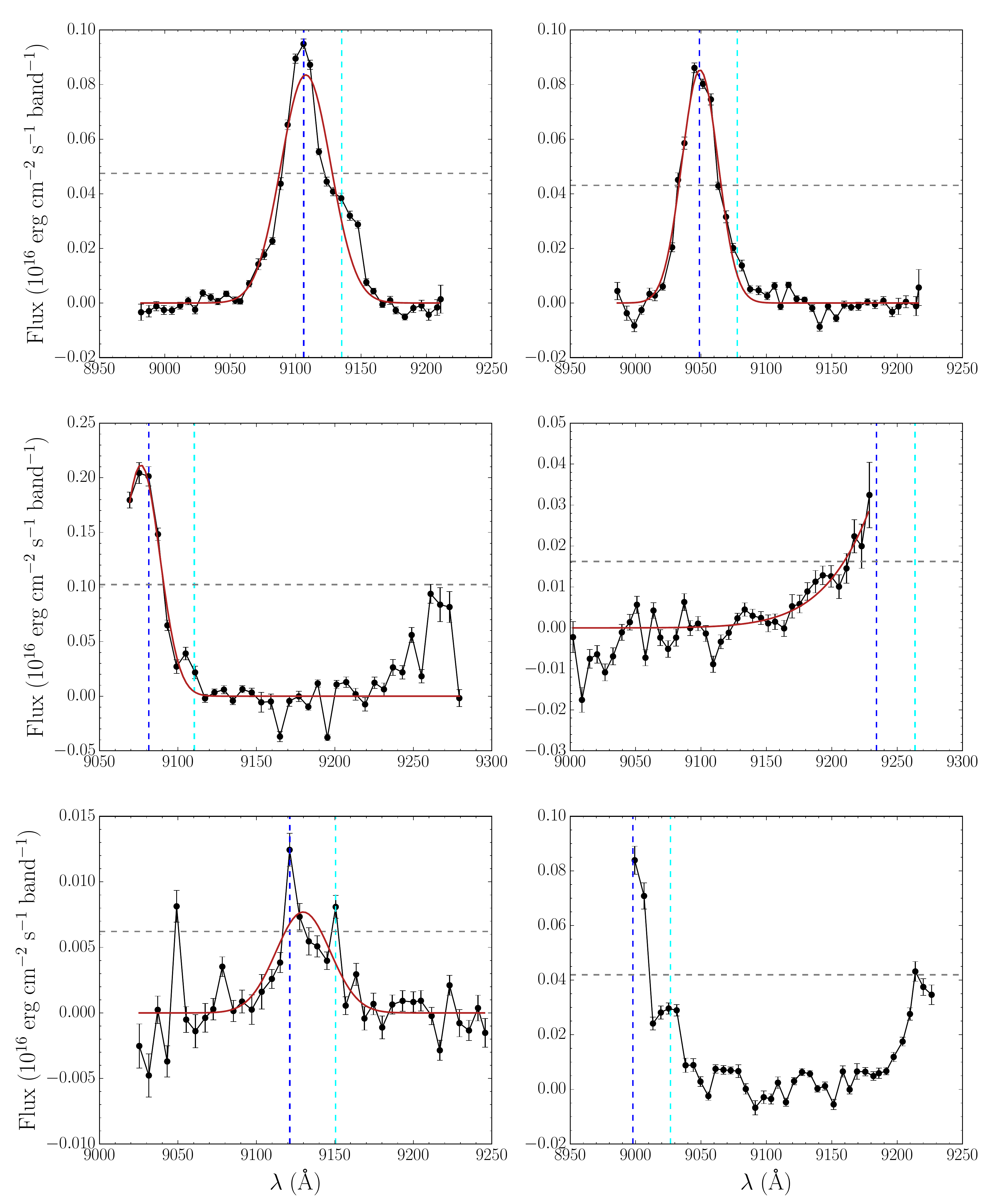}
\captionof{figure}{Objects from OTELO survey showing  H$\alpha$+[NII] emission and classified as BLAGNs. The red lines show the best fit to a Gaussian profile, while the grey dashed lines represent half the maximum value of the pseudo-spectra. The blue and cyan vertical lines mark the positions of the  H$\alpha$ and [NII] emission lines, respectively. In the last case, no fitting could be made due to the truncation of the line. However, the object was included in the final sample because its width is comparable to the rest of the objects selected as BLAGNs.}
\label{BLAGN}
\end{minipage}

\section{AGN morphologies (examples)}\label{appendix_morphology}

\vspace{20pt}

\begin{figure}[!h]
\vspace{75pt}
\centering
\includegraphics[width=0.8\textwidth]{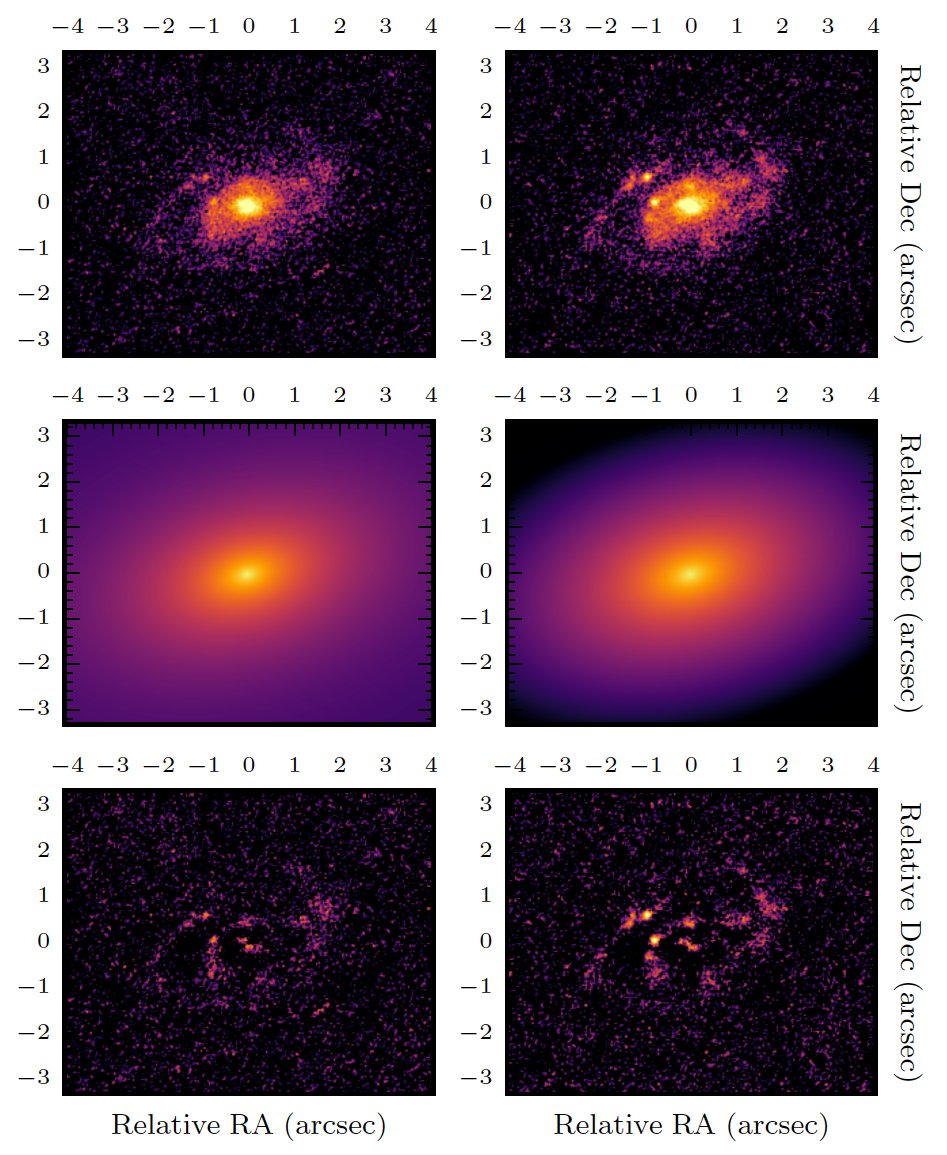}
\caption[Morphology of the object \#1873]{Morphology of the object \#1873. This object was classified as a BLAGNs at $z\sim 0.4$ and is clearly a spiral galaxy (S, late-type). The spiral arms are visible in the residual. First row: HST images of the object. Second row: GALFIT models. Third row: GALFIT residuals obtained by subtracting its model from the original image. The first column shows the V filter of HST (F814W) and the second one the I filter (F606W).}
\label{1873}
\end{figure}

\begin{figure}[!h]
\vspace{65pt}
\centering
\includegraphics[width=0.55\textwidth]{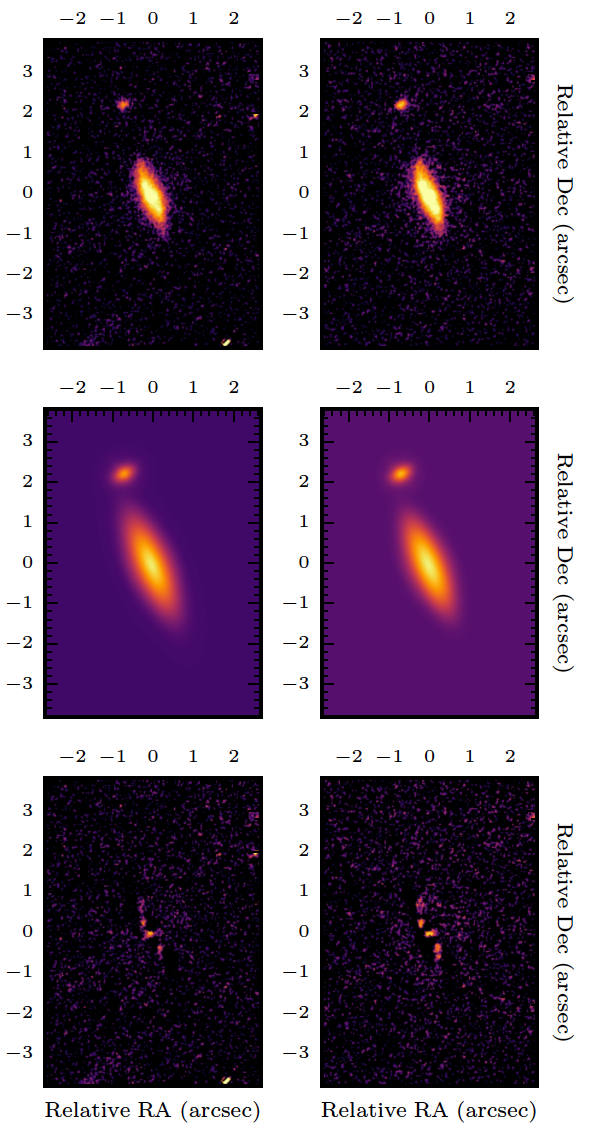}
\caption[Morphology of the object \#8762]{Morphology of the object \#8762. This object is a BLAGN at $z\sim 0.4$. It is a barred spiral galaxy (SB), whose bar appears visible in the residual images. First row: HST images of the object. Second row: GALFIT models. Third row: GALFIT residuals obtained by subtracting its model to the original image. The first column shows the V filter of HST (F814W) and the second one the I filter (F606W)..}
\label{8762}
\end{figure}

\begin{figure}[!h]
\vspace{75pt}
\centering
\includegraphics[width=0.8\textwidth]{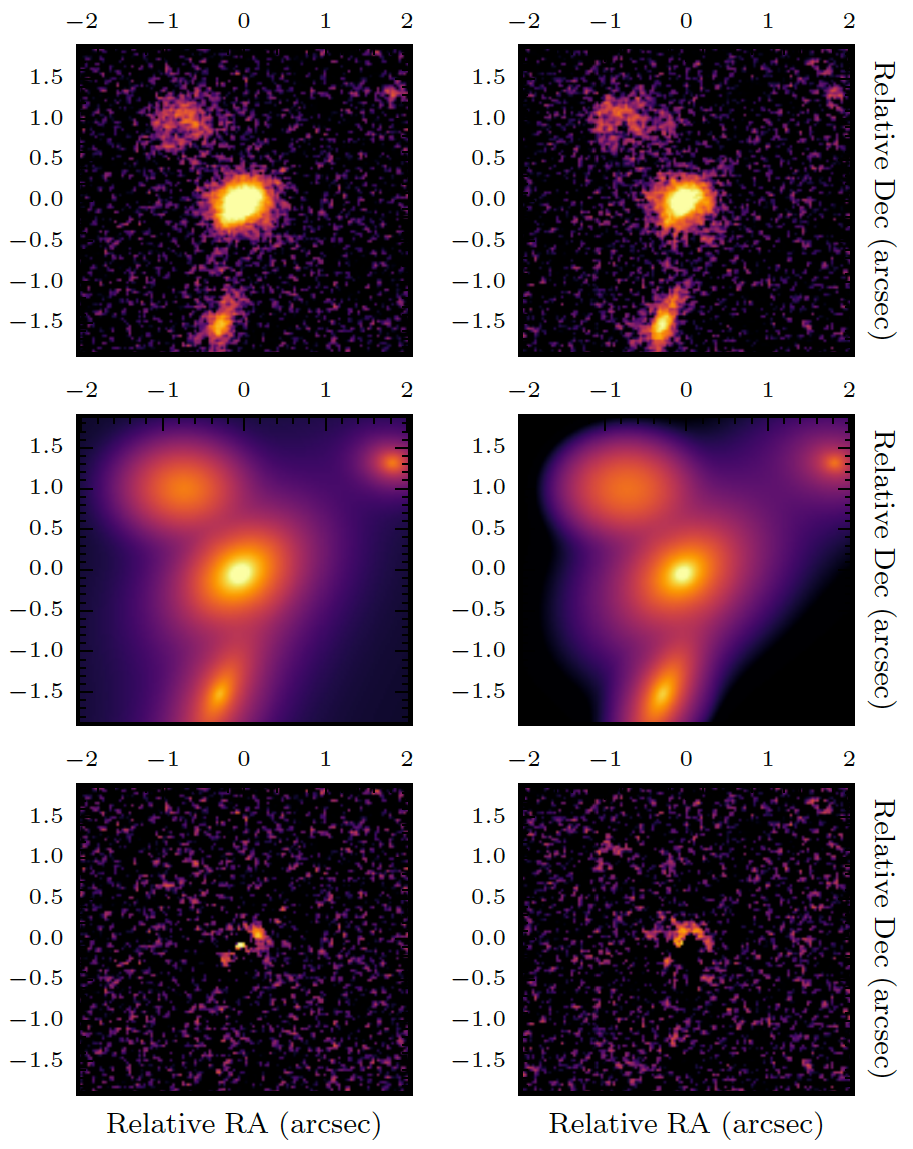}
\caption[Morphology of the object \#2146]{Morphology of the object \#2146. This object is a BLAGN at $z\sim 0.4$. It was classified as a late-type and flagged as a multiple object. First row: HST images of the object. Second row: GALFIT models. Third row: GALFIT residuals obtained by subtracting its model to the original image. The first column shows the V filter of HST (F814W) and the second one the I filter (F606W)..}
\label{2146}
\end{figure}

\begin{figure}[!h]
\vspace{75pt}
\centering
\includegraphics[width=0.60\textwidth]{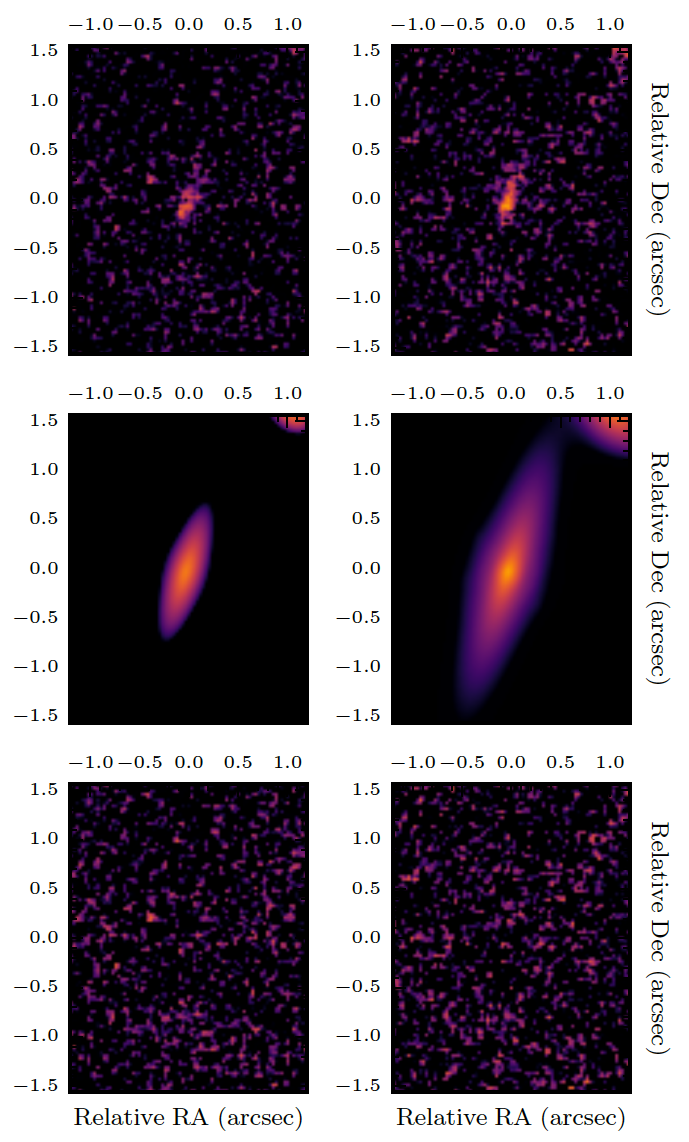}
\caption[Morphology of the object \#3854]{Morphology of the object \#3854.  This object was classified as a NLAGN at $z \sim 0.4$ and was not assigned any morphological type (unclassifiable). First row: HST images of the object. Second row: GALFIT models. Third row: GALFIT residuals obtained by subtracting its model to the original image. The first column shows the V filter of HST (F814W) and the second one the I filter (F606W).}
\label{3854}
\end{figure}

\begin{figure}[!h]
\vspace{75pt}
\centering
\includegraphics[width=0.70\textwidth]{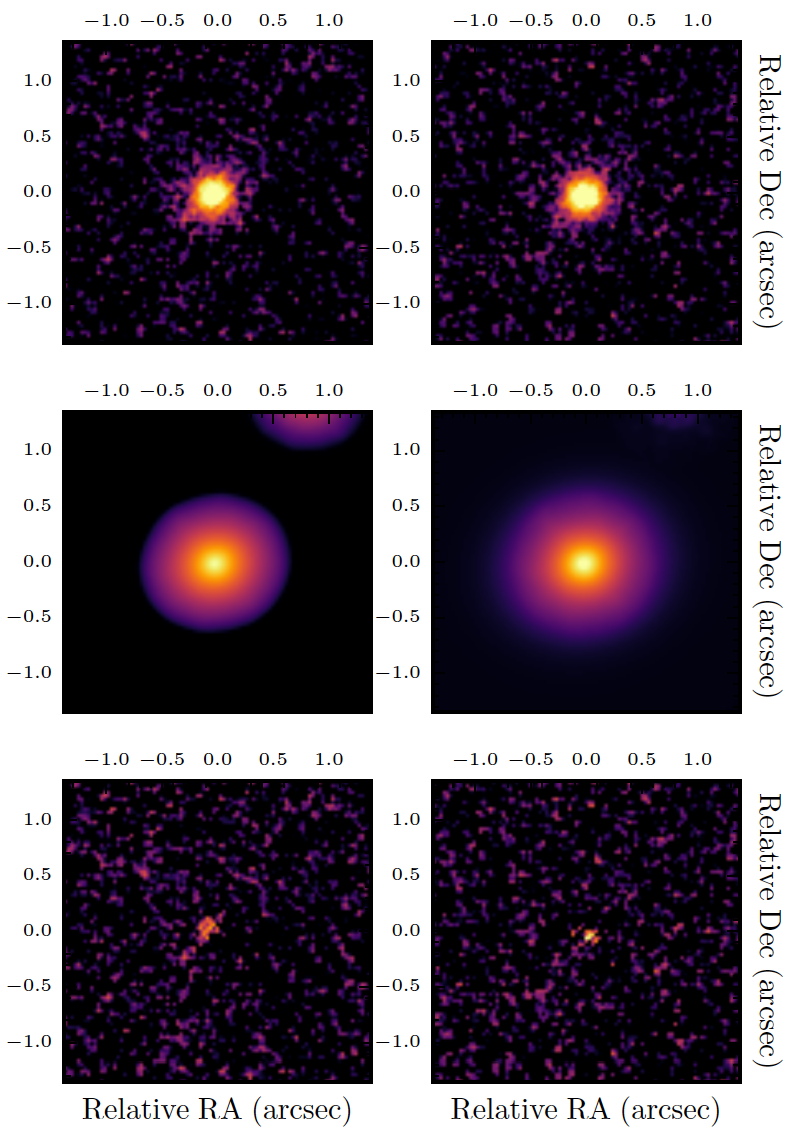}
\caption[Morphology of the object \#6395]{Morphology of the object \#6395. This object was classified as a NLAGN at $z\sim 0.4$ with a spheroidal morphology (early-type). First row: HST images of the object. Second row: GALFIT models. Third row: GALFIT residuals obtained by subtracting its model to the original image. The first column shows the V filter of HST (F814W) and the second one the I filter (F606W).}
\label{6395}
\end{figure}

\begin{figure}[!h]
\vspace{40pt}
\centering
\includegraphics[width=0.6\textwidth]{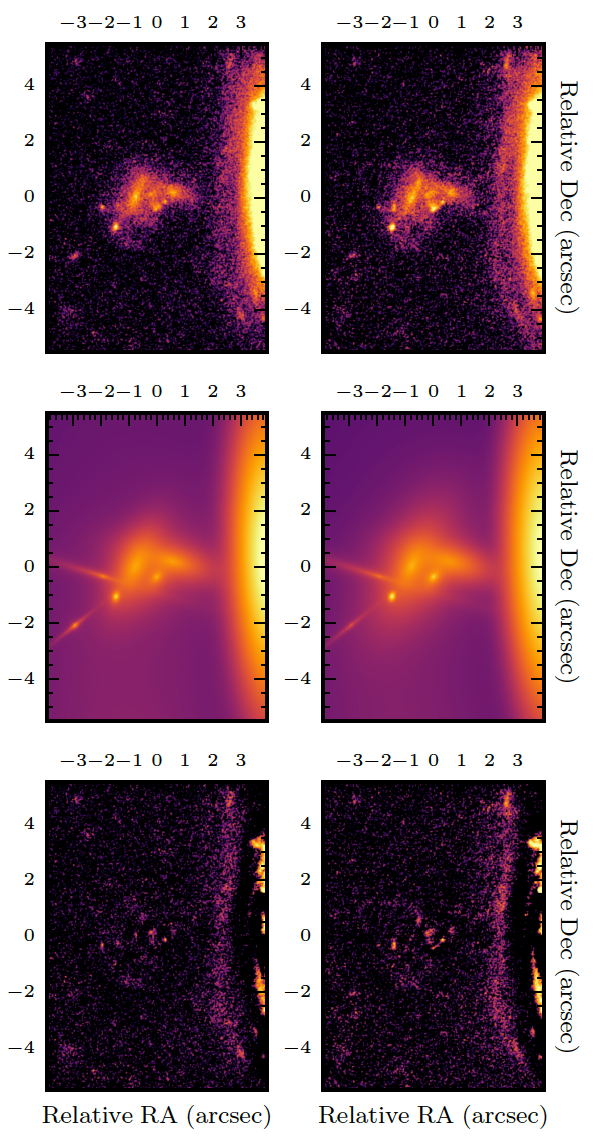}
\caption[Morphology of the object \#5662]{Morphology of the object \#5662. This object, selected as an AGN by its X-rays emission, is part of a system with multiple interacting components. Due to its complexity, it was not assigned any morphological type (unclassifiable). First row: HST images of the object. Second row: GALFIT models. Third row: GALFIT residuals obtained by subtracting its model to the original image. The first column shows the V filter of HST (F814W) and the second one the I filter (F606W).}
\label{5662}
\end{figure}

\begin{figure}[!h]
\vspace{60pt}
\centering
\includegraphics[width=0.50\textwidth]{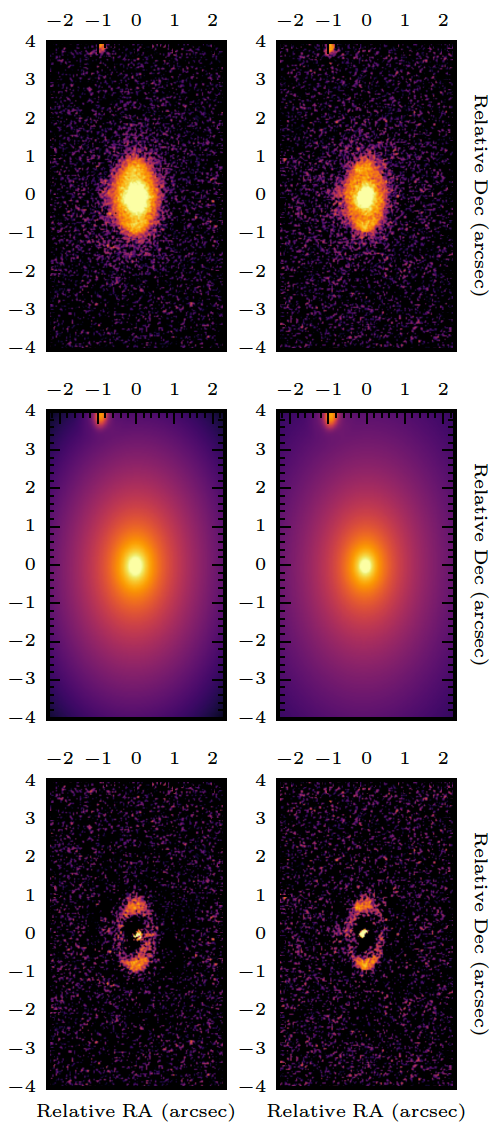}
\caption[Morphology of the object \#3216]{Morphology of the object \#3216. This object is an X-ray-selected AGN. It is a face-on spiral, whose arms are visible in the residual images (late-type). First row: HST images of the object. Second row: GALFIT models. Third row: GALFIT residuals obtained by subtracting its model to the original image. The first column shows the V filter of HST (F814W) and the second one the I filter (F606W).}
\label{3216}
\end{figure}

\begin{figure}[!h]
\vspace{75pt}
\centering
\includegraphics[width=0.70\textwidth]{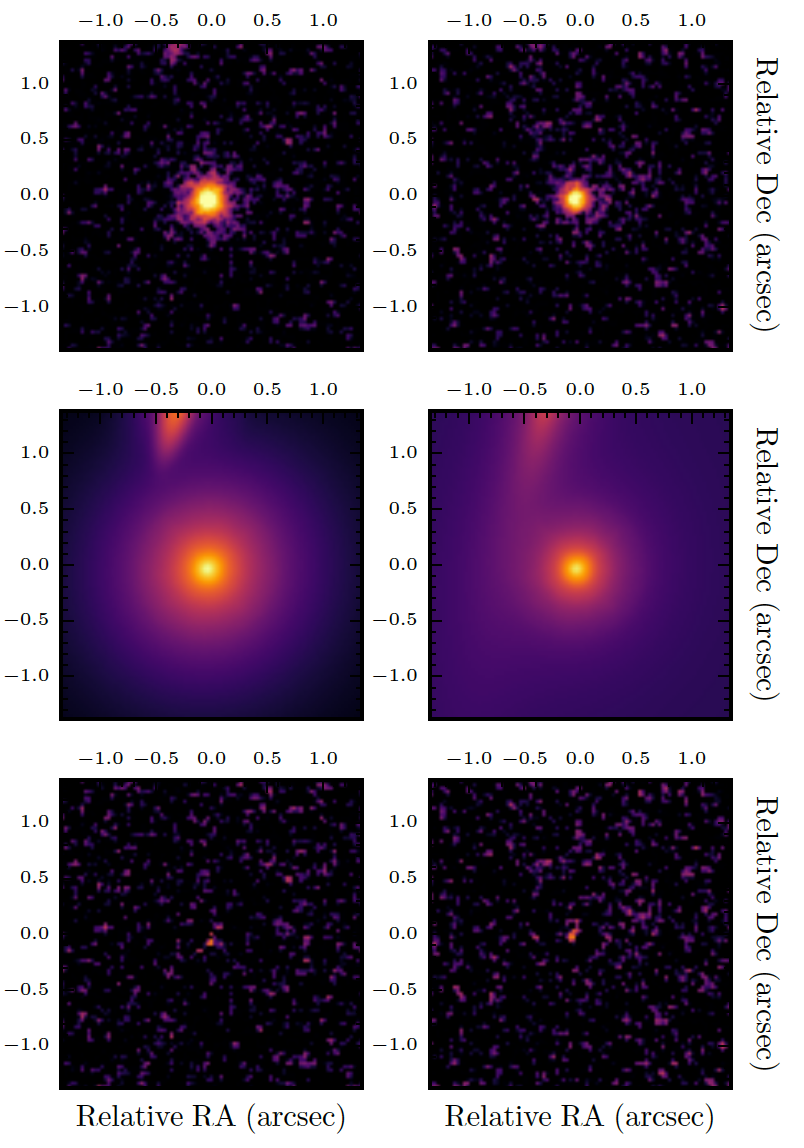}
\caption[Morphology of the object \#6173]{Morphology of the object \#6173. This object is an X-ray-selected AGN. It has a point-like appearance, and is probably a QSO, just like the object \#8351 shown in Fig. \ref{8351}. First row: HST images of the object. Second row: GALFIT models. Third row: GALFIT residuals obtained by subtracting its model to the original image. The first column shows the V filter of HST (F814W) and the second one the I filter (F606W).}
\label{6173}
\end{figure}

\begin{figure}[!h]
\vspace{75pt}
\centering
\includegraphics[width=0.70\textwidth]{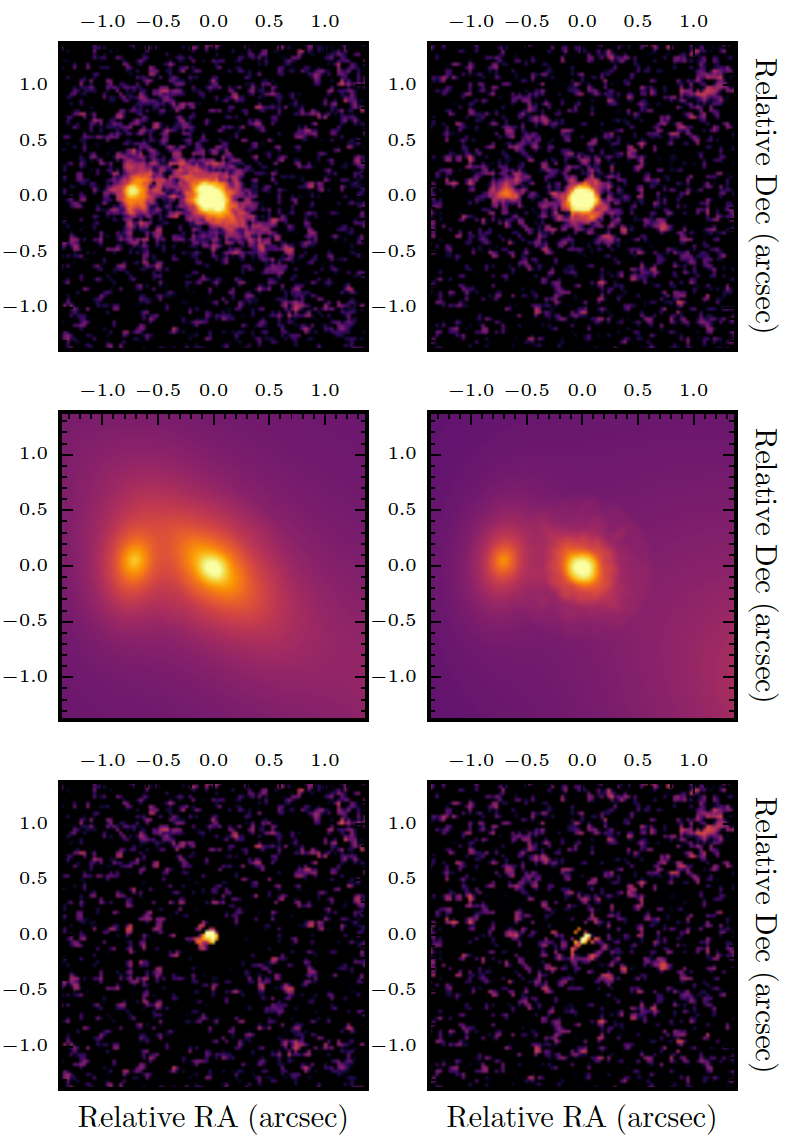}
\caption[Morphology of the object \#5495]{Morphology of the object \#5495. This object is an X-ray-selected AGN. It seems to be an early-type galaxy (probably lenticular, S0) with a companion. First row: HST images of the object. Second row: GALFIT models. Third row: GALFIT residuals obtained by subtracting its model to the original image. The first column shows the V filter of HST (F814W) and the second one the I filter (F606W).}
\label{5495}
\end{figure}

\begin{figure}[!h]
\vspace{75pt}
\centering
\includegraphics[width=0.55\textwidth]{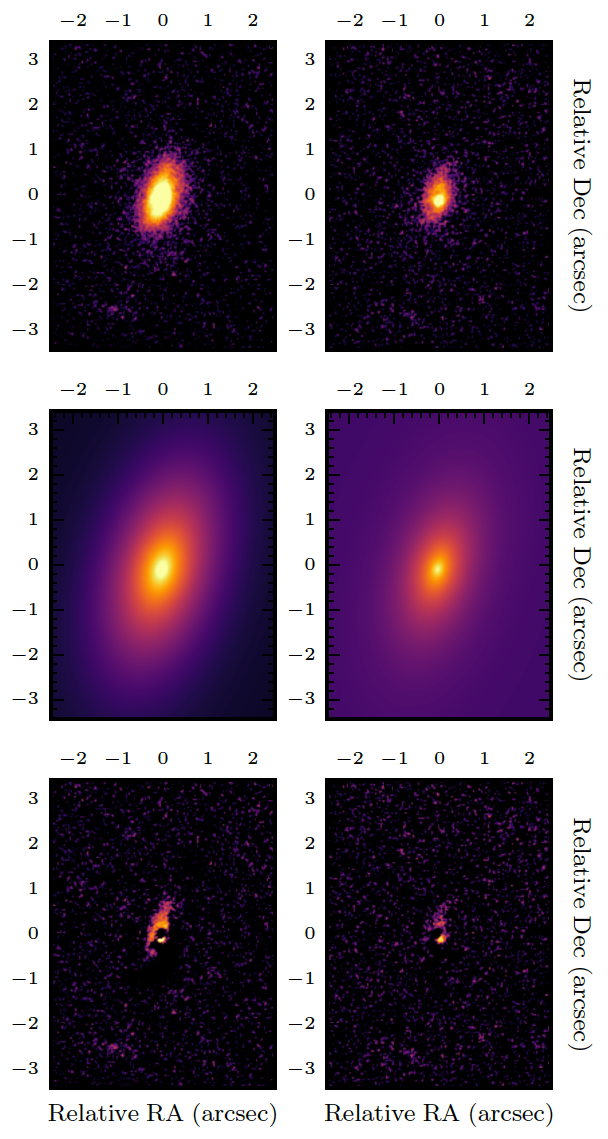}
\caption[Morphology of the object \#8459]{Morphology of the object \#8459.  This object is an unobscured AGN (selected both in X-rays and IR) with a morphological disc type (late-type). First row: HST images of the object. Second row: GALFIT models. Third row: GALFIT residuals obtained by subtracting its model to the original image. The first column shows the V filter of HST (F814W) and the second one the I filter (F606W).}
\label{8459}
\end{figure}

\begin{figure}[!h]
\vspace{75pt}
\centering
\includegraphics[width=0.70\textwidth]{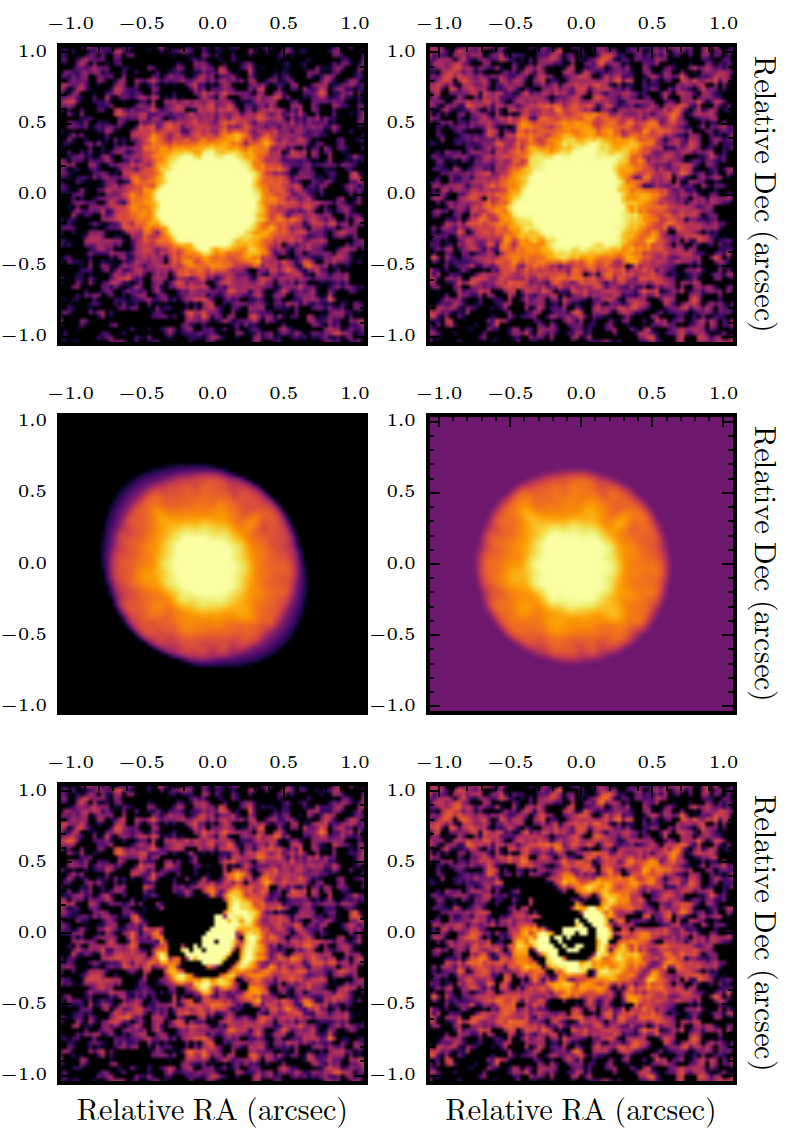}
\caption[Morphology of the object \#8351]{Morphology of the object \#8351. This object was selected as an AGN both in X-rays and MIR. It has a point-like appearance, and is probably a QSO, just like the object \#6173 shown in Fig. \ref{6173}.  First row: HST images of the object. Second row: GALFIT models. Third row: GALFIT residuals obtained by subtracting its model to the original image. The first column shows the V filter of HST (F814W) and the second one the I filter (F606W).}
\label{8351}
\end{figure}

\begin{figure}[!h]
\vspace{80pt}
\centering
\includegraphics[width=1\textwidth]{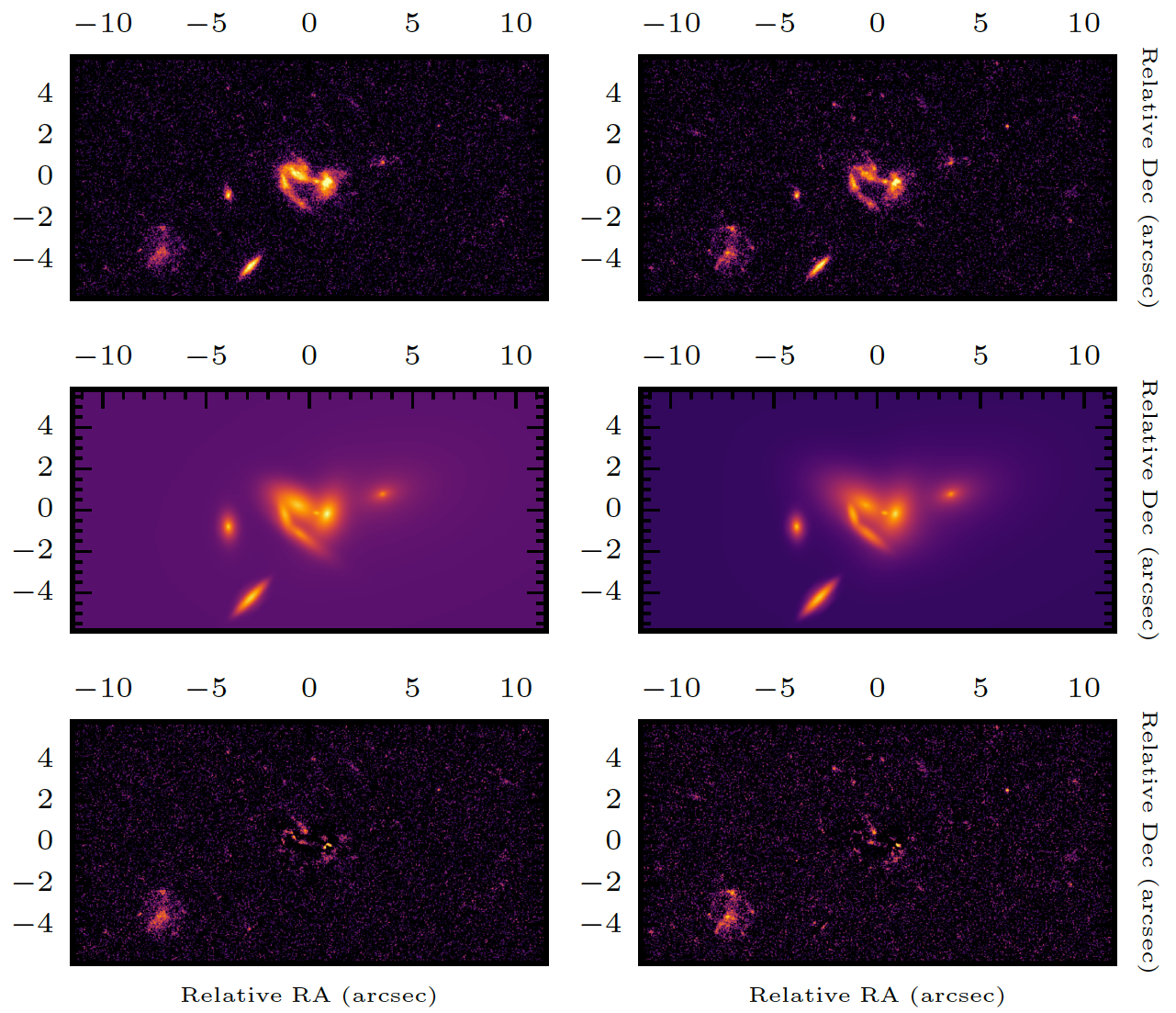}
\caption[Morphology of the object \#11168]{Morphology of the object \#11168. This system was selected by IR methods as an AGN. Two interacting spiral galaxies are visible. First row: HST images of the object. Second row: GALFIT models. Third row: GALFIT residuals obtained by subtracting its model to the original image. The first column shows the V filter of HST (F814W) and the second one the I filter (F606W).}
\label{11168}
\end{figure}

\begin{figure}[!h]
\vspace{75pt}
\centering
\includegraphics[width=0.8\textwidth]{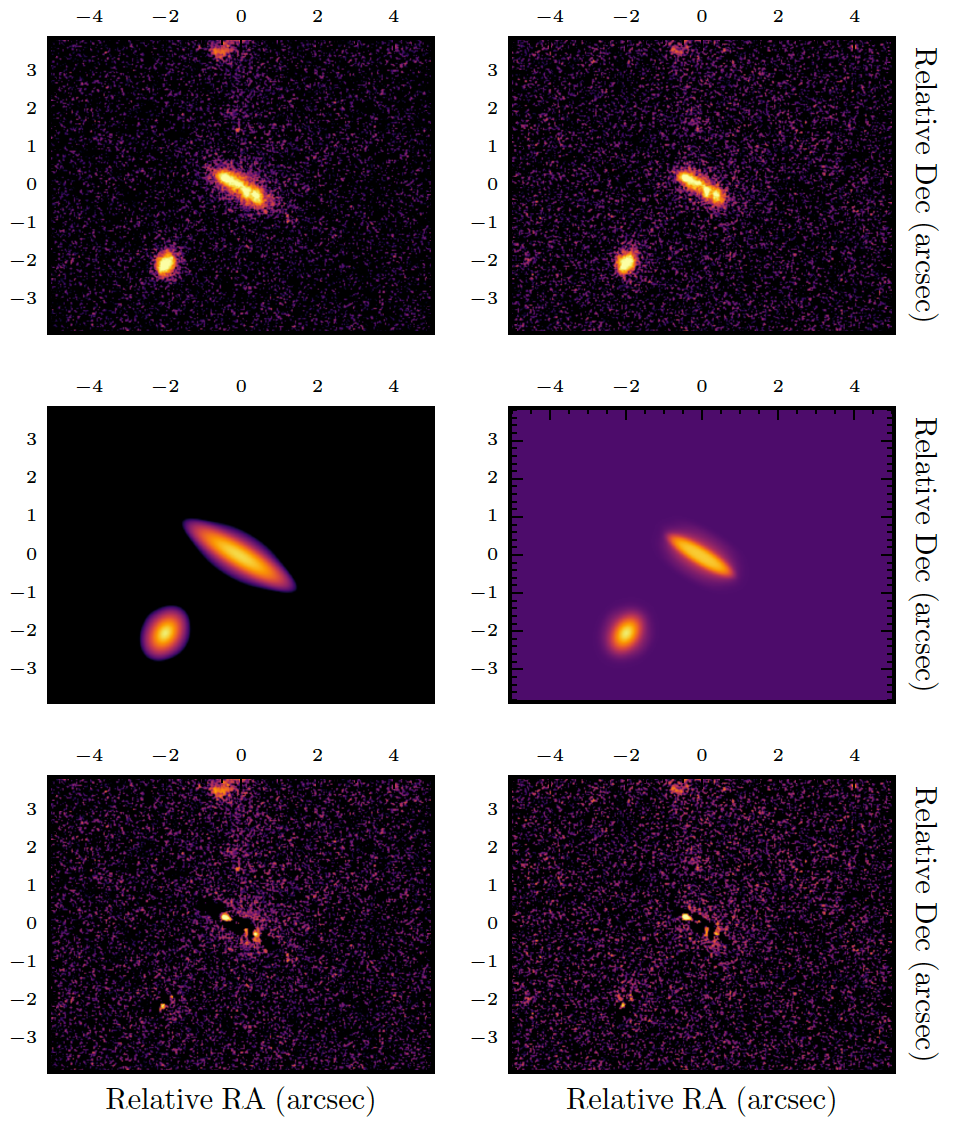}
\caption[Morphology of the object \#7800]{Morphology of the object \#7800. This object is an obscured AGN. It was selected with IR methods and classified as an irregular object (late-type), seemingly a chain-galaxy. First row: HST images of the object. Second row: GALFIT models. Third row: GALFIT residuals obtained by subtracting its model to the original image. The first column shows the V filter of HST (F814W) and the second one the I filter (F606W).}
\label{7800}
\end{figure}

\begin{figure}[!h]
\vspace{60pt}
\centering
\includegraphics[width=0.65\textwidth]{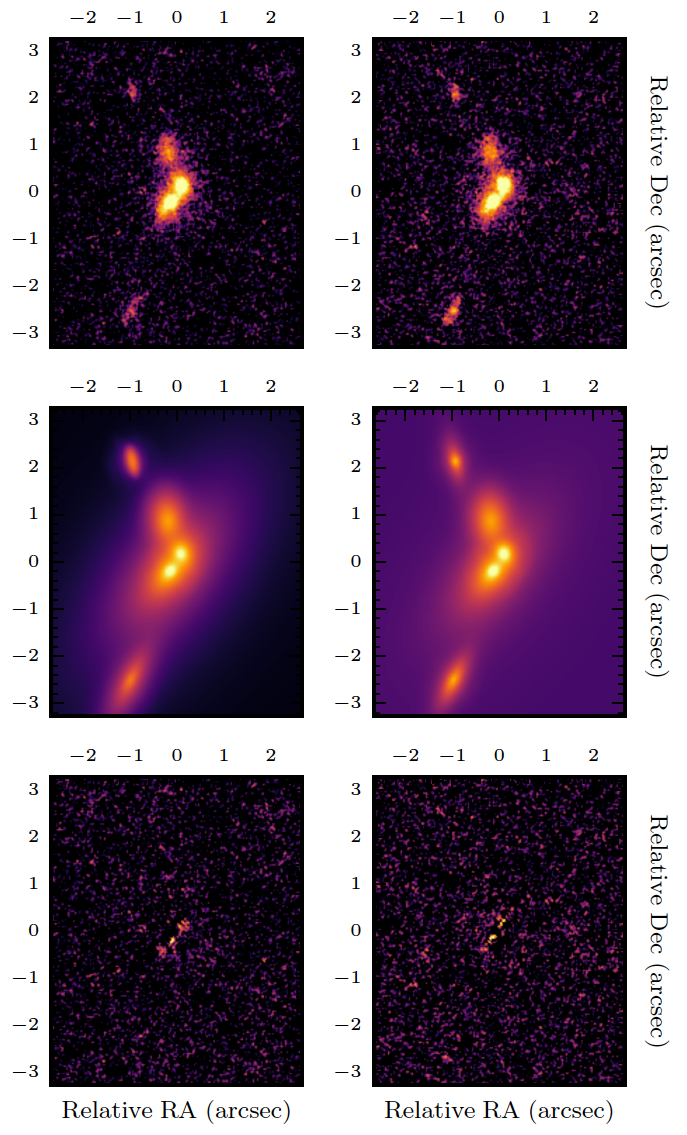}
\caption[Morphology of the object \#10965]{Morphology of the object \#10965. This object is an obscured AGN, selected only by IR methods, which seems to be interacting. It was classified as a late-type galaxy. First row: HST images of the object. Second row: GALFIT models. Third row: GALFIT residuals obtained by subtracting its model to the original image. The first column shows the V filter of HST (F814W) and the second one the I filter (F606W).}
\label{10965}
\end{figure}

\begin{figure}[!h]
\vspace{75pt}
\centering
\includegraphics[width=0.60\textwidth]{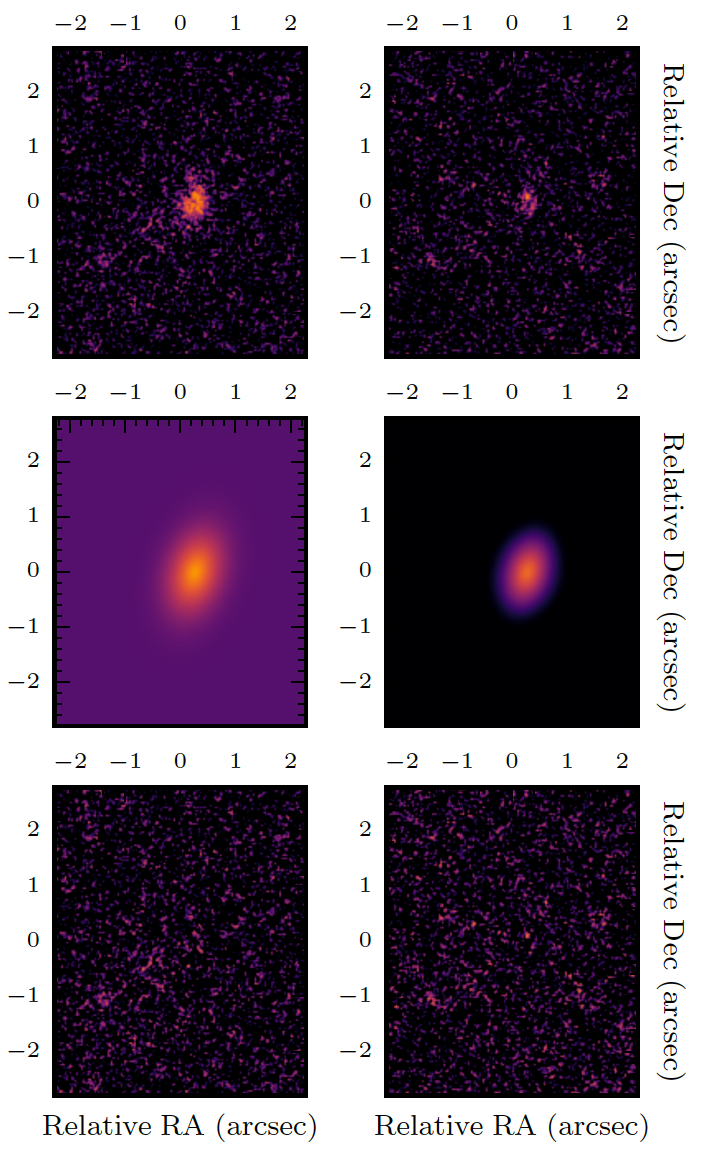}
\caption[Morphology of the object \#7772]{Morphology of the object \#7772. This object is a faint obscured AGN, classified as an irregular galaxy (late-type). First row: HST images of the object. Second row: GALFIT models. Third row: GALFIT residuals obtained by subtracting its model to the original image. The first column shows the V filter of HST (F814W) and the second one the I filter (F606W).}
\label{7772}
\end{figure}

In the following pages of this appendix, the morphologies of some of the AGNs found in OTELO are shown as representative examples:

\begin{itemize}
\item Broad-line AGNs at $z\sim 0.4$: Figures \ref{1873}, \ref{8762} and \ref{2146}.
\item Narrow-line AGNs at $z\sim 0.4$: Figures \ref{3854} and \ref{6395}.
\item X-rays selected AGNs: Figures \ref{5662}-\ref{5495}.
\item Unobscured AGNs (MIR+X-rays selection): Figure \ref{8459} and \ref{8351}.
\item Obscured AGNs (MIR selection only): Figures \ref{11168}-\ref{7772}.
\end{itemize}

The figures display the object in the original HST images along with their GALFIT model and residual, both described in section \ref{section_morphology}.

\newpage

\newpage

\newpage

\newpage

\newpage

\newpage

\newpage

\newpage

\newpage

\newpage

\newpage

\newpage

\newpage

\newpage

\newpage

\end{appendix}

\end{document}